\documentclass{pasj00}
\color{black}
\tenpoint

\SetRunningHead{K. Sato et al.}{
Suzaku observation of NGC~507}

\title{
Suzaku Observation of Group of Galaxies NGC~507:\\
Temperature and Metal Distributions \\
in the Intra-cluster Medium}
\author{
 Kosuke \textsc{Sato},\altaffilmark{1,2}
 Kyoko \textsc{Matsushita},\altaffilmark{1}
 Yoshitaka \textsc{Ishisaki},\altaffilmark{3} 
 Noriko \textsc{Y.~Yamasaki},\altaffilmark{4}\\
 Manabu \textsc{Ishida},\altaffilmark{4} 
and Takaya \textsc{Ohashi},\altaffilmark{3} 
}
\altaffiltext{1}{
Department of Physics, Tokyo University of Science,
1-3 Kagurazaka, Shinjuku-ku, Tokyo 162-8601}
\altaffiltext{2}{
Graduate School of Natural Science and Technology,
Kanazawa University, 
Kakuma, Kanazawa, Ishikawa 920-1192}
\email{ksato@astro.s.kanazawa-u.ac.jp}
\altaffiltext{3}{
Department of Physics, Tokyo Metropolitan University,
1-1 Minami-Osawa, Hachioji, Tokyo 192-0397}
\altaffiltext{4}{
Institute of Space and Astronautical Science (ISAS),
Japan Aerospace Exploration Agency, \\
3-1-1 Yoshinodai, Sagamihara, Kanagawa 229-8510}
\Received{2008~March~10}
\Accepted{2008~May~14}
\Published{---}

\begin{document}
\maketitle

\begin{abstract}
Temperature and abundance distributions of the intra-cluster
medium (ICM) in the NGC~507 group of galaxies were studied with Suzaku.
Observed concentric annular spectra were well-represented by a two
temperature model for ICM, and we found steeper abundance gradients
for Mg, Si, S, and Fe compared with O in the central region.
Abundance ratios of $\alpha$-elements to iron were found to be
similar to those in other groups and poor clusters.  We
calculated metal mass-to-light ratios for Fe, O and Mg (IMLR, OMLR,
MMLR) for NGC~507, and values for different systems were compared.  Hotter
and richer systems tend to show higher values of IMLR, OMLR, and MMLR\@.  OMLR
and MMLR were measured to an outer region for the first time with
Suzaku, while IMLR was consistent with that with ASCA\@. We also
looked into 2-dimensional map of the hardness ratio, but found
no significant deviation from the circular symmetry.
\end{abstract}

\section{Introduction}

Recent X-ray observations have shown metal abundance profiles in the
ICM based on the spatially resolved spectra.  Groups of galaxies carry
important information about the entire history of cosmic chemical
evolution, since the groups contain 50--70\% of galaxies in the
universe (e.g.\ \cite{mulchaey06}) and they serve as building blocks
of clusters of galaxies.  In order to know how the ICM has been
enriched, we need to measure the amount and distribution of metals in
the ICM\@.  Because Si and Fe are both synthesized in type Ia and type
II supernovae (SNe Ia and II), we need to know O and Mg abundances,
which are synthesized practically in SNe II, in resolving the past
metal enrichment process in ICM by supernovae in early-type galaxies
\citep{arnaud92,renzini93}.  Comparison of the ICM properties between
groups and clusters of galaxies will also enable us to understand the
bottom-up process of the hierarchical structure formation in the
universe.

ASCA measured the distribution of the heavy elements, such as Si and
Fe, in the ICM
\citep{fukazawa98,fukazawa00,finoguenov00,finoguenov01}.
\citet{renzini97} and \citet{makishima01} summarized
iron-mass-to-light ratios (IMLR) for various objects, as a function of
their plasma temperature serving as a measure of the system richness
with ASCA, and they showed that the early-type galaxies released large
amount of metals formed through past supernovae explosions as shown
earlier by \citet{arnaud92}.  Recent XMM-Newton and Chandra
observations have enabled us to study properties of the heavy elements
in the ICM in detail.  These observations showed not only Si and Fe,
but also O and Mg abundance profiles, however the O and Mg abundance
measurements have been limited only for the central regions of very
bright clusters or groups of galaxies dominated by cD galaxies in a
reliable manner
\citep{finoguenov02,xu02,matsushita03,tamura03,buote03a,buote03b,humphrey06}.
\citet{tamura04} derived IMLR for five clusters, and the oxygen mass
for several clusters with XMM-Newton.  However, oxygen-mass-to-light
ratios (OMLR) for rich clusters are not reliable due to the lower
emissivity of O\emissiontype{VII} and O\emissiontype{VIII} lines in
high temperatures.  \citet{matsushita07b} showed different
distribution profiles among the metals, which suggest difference in
the process of metal enrichment.  The abundance measurements of O and
Mg with XMM-Newton, particularly for the outer regions of groups and
clusters, are quite difficult due to the relatively high intrinsic
background. Suzaku XIS can measure all the main elements from O to Fe,
because it realizes lower background and higher spectral sensitivity,
especially below 1 keV \citep{koyama07}.  \citet{matsushita07a},
\citet{sato07a},\citet{sato08}, and \citet{tokoi08} have shown the
abundance profiles of O, Mg, Si, S, and Fe with Suzaku to the outer
regions with good accuracy.

NGC~507 is a nearby group of galaxies ($z=0.01646$) characterized by a
smooth distribution of ICM\@.  
The ICM properties have been studied with ROSAT 
\citep{kim95,paolillo03}, ASCA \citep{matsumoto97}, 
Chandra \citep{kraft04,humphrey06,rasmussen07}, and XMM-Newton \citep{kim04}.
\citet{kim95} revealed a cooler central region with ROSAT PSPC, 
and \citet{kim04} showed the supersolar metal abundances 
within the $D_{25}$ ellipse of NGC~507 with XMM-Newton. 
On the other hand, \citet{humphrey06} showed a near solar metal abundances, 
the same as derived by \citet{kraft04} with Chandra. 
\citet{kim04} and \citet{humphrey06} showed the $\alpha$-elements to 
iron ratios, and both the abundance ratios were almost consistent 
to be $\sim 1$ solar.  
\citet{kraft04} also reported a sharp edge or a discontinuity 
in the radial surface brightness 
profile 55 kpc east and southeast of NGC~507 
covering an $\sim125^{\circ}$ arc. NGC~507 is also a known radio source 
and has been classified as an FR I radio galaxy  \citep{fanti86}, and
the edge of the radio lobe corresponds to the discontinuity of 
the X-ray surface brightness as shown in \citet{kraft04}.

This paper reports on results from Suzaku observations of NGC~507
out to $13'\simeq 260\; h_{70}^{-1}$~kpc, corresponding to
$\sim 0.24\; r_{180}$.  We use $H_0=70$
km~s$^{-1}$~Mpc$^{-1}$, $\Omega_{\Lambda} = 1-\Omega_M = 0.73$ in this
paper.  At a redshift of $z=0.01646$, $1'$ corresponds to 20.1~kpc,
and the virial radius, $r_{\rm 180} = 1.95\;
h_{100}^{-1}\sqrt{k\langle T\rangle/10~{\rm keV}}$~Mpc
\citep{markevitch98}, is 1.08~Mpc ($54'$) for an average temperature
of $k\langle T\rangle = 1.5$~keV\@.  Throughout this paper we adopt
the Galactic hydrogen column density of $N_{\rm H} = 5.24\times
10^{20}$ cm$^{-2}$ \citep{dickey90} in the direction of NGC~507\@.
Unless noted otherwise, the solar abundance table is given by
\citet{anders89}, and the errors are in the 90\% confidence region 
for a single interesting parameter.

\begin{table*}
\caption{Suzaku Observation logs for NGC~507.}
\label{tab:1}
\begin{tabular}{lccccc} \hline 
Object & Seq. No. & Obs. date & \multicolumn{1}{c}{(RA, Dec)$^\ast$} &Exp.&After screening \\
&&&J2000& ksec &(BI/FI) ksec \\
\hline 
NGC~507 & 801017010 & 2006-07-28T14:51:16 & (\timeform{01h23m40.0s}, \timeform{+33D15'21''})& 79.6& 79.2/79.2\\
\hline\\[-1ex]
\multicolumn{6}{l}{\parbox{0.9\textwidth}{\footnotesize 
\footnotemark[$\ast$]
Average pointing direction of the XIS, written in the 
RA\_NOM and DEC\_NOM keywords of the event FITS files.}}\\
\end{tabular}
\end{table*}

\begin{figure}
\begin{center}
\begin{minipage}{0.45\textwidth}
\centerline{
\FigureFile(\textwidth,\textwidth){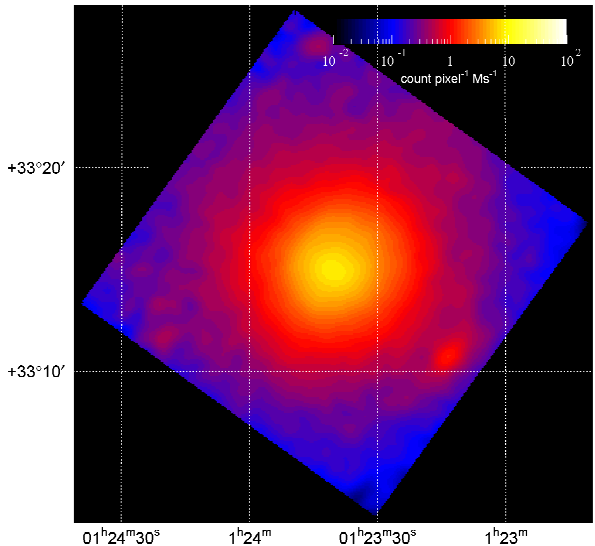}}
\caption{
Combined XIS image of central observation in the
0.5--4.0 keV energy range. The observed XIS0-3 images were added on the sky
coordinate after removing each calibration source region,
and smoothed with $\sigma=16$ pixel $\simeq 17''$ Gaussian.
Estimated components of extragalactic X-ray background (CXB)
and instrumental background (NXB) were subtracted,
and the exposure was corrected, though vignetting was not corrected.
}\label{fig:1}
\end{minipage}
\end{center}
\end{figure}

\section{Observations and Data Reduction}\label{sec:obs}

\subsection{Observation}
\label{subsec:obs}

Suzaku observed the central region of NGC~507 in July 2006
(PI:~K. Sato) with an exposure time of 79.6 ks.  The observation log
is given in table~\ref{tab:1}, and the XIS image in 0.5--4~keV is
shown in figure~\ref{fig:1}.  We analyze only the XIS data in this
paper, although Suzaku observed the object with both XIS and HXD, 
which acquired the data.  The XIS instrument consists of four sets 
of X-ray CCDs (XIS~0, 1, 2, and 3). XIS~1 is a back-illuminated (BI) 
sensor, while XIS~0, 2, and 3 are front-illuminated (FI)\@.  The 
instrument was operated in the Normal clocking mode (8~s exposure 
per frame), with the standard $5\times 5$ or $3\times 3$ editing mode.

It is known that the optical blocking filters (OBF) of the XIS have
gradually been contaminated by outgassing from the satellite.  The
thickness of the contaminant is different among the sensors, and is
also dependent on the location on the CCD chips in the way that 
it is thickest in the center.  Estimated column
densities (C/O=6 in number ratio is assumed) during the observation at
the center of the CCD are given in 
table~\ref{tab:2}.\footnote{ Calibration database file of {\tt ae\_xi{\it
N}\_contami\_20061024.fits} was used for the estimation of the XIS
contamination ($N=0,1,2,3$ denotes the XIS sensor).}  We included
these effects in the calculation of the the Ancillary Response File
(ARF) by the ``xissimarfgen'' ftools task of 2006-10-26 version
\citep{ishisaki07}.  Since the energy resolution also slowly degraded
after the launch, due to radiation damage, this effect was included
in the Redistribution Matrix File (RMF) by the ``xisrmfgen'' Ftools
task of the 2006-10-26 version.

\begin{table}
\caption{Estimated column density of the contaminant for each sensor at 
the center of CCD in units of 10$^{18}$~cm$^{-2}$.
}\label{tab:2}
\begin{center}
\begin{tabular}{cllll} 
\hline
\hspace*{8em} &XIS0 & XIS1 & XIS2 &XIS3\\
\hline
Carbon $\dotfill$ & 2.52   & 3.94  & 3.83 & 5.78 \\
Oxygen $\dotfill$ & 0.419  & 0.657 & 0.638 & 0.963 \\
\hline 
\end{tabular}
\end{center}
\end{table}

\subsection{Data Reduction}

We used version 1.2 processing data \citep{mitsuda07}, and the
analysis was performed with HEAsoft version 6.1.1 and XSPEC 11.3.2t.
The analysis method was almost the same as those in \citet{sato07a} and
\citet{sato08}.  However, because the observations were not supported
by the Good-Time Intervals (GTI) defined to exclude the telemetry
saturation by the XIS team, we could not execute the GTI correction.  
The light curve of each sensor in the 0.3--10~keV range with a 16~s time
bin was also examined in order to exclude periods with anomalous event
rates which were greater or less than $\pm 3\sigma$ around the mean.  After the
above screenings, the remaining exposure time of the observation stayed
almost unchanged as shown in table~\ref{tab:1}.  The exposure after
the screening was essentially the same as that before screening in
table~\ref{tab:1}, which indicates that the non X-ray background (NXB) was
almost stable during the observation.  Event screening with cut-off
rigidity (COR) was not performed in our data.

In order to subtract the NXB and the extra-galactic
cosmic X-ray background (CXB), we employed the dark Earth
database of 770~ks exposure, provided by the XIS team for the NXB, and
employed the CXB spectrum given by \citet{kushino02}.  These analysis
methods were also the same as in \citet{sato07a} and \citet{sato08}.
In order to remove influence from bright sources, 
we eliminated the regions within 80 arcsec from 
(\timeform{01h23m14.0s}, \timeform{+33D10'57''}) and 
within 45 arcsec from (\timeform{01h23m39.1s}, \timeform{+33D09'04''}) 
in our field of view. However, because the spatial resolution of 
Suzaku is not enough to resolve all point sources especially at the 
central region, effect from the remaining weaker
point sources was taken into account in the spectral fitting function 
in the next sections.

\begin{figure}
\centerline{\FigureFile(0.45\textwidth,8cm){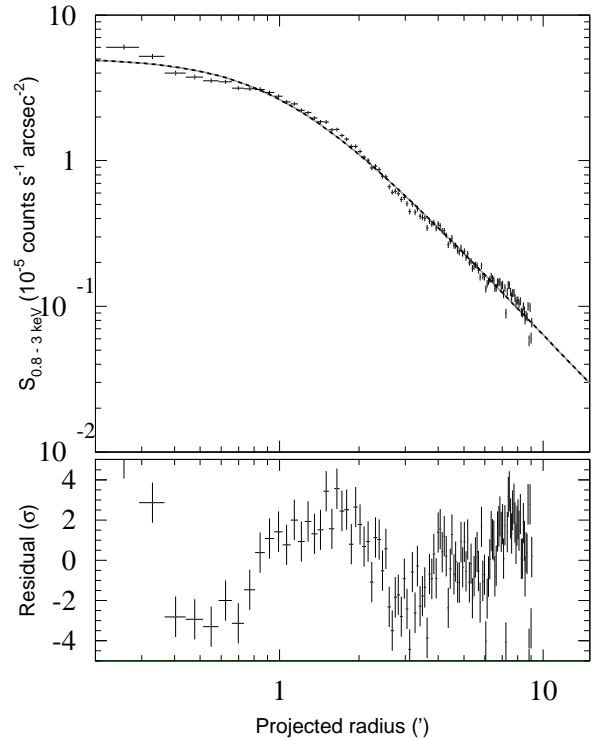}}
\vspace*{-1ex}
\caption{
In the upper panel, a radial profile of the surface brightness 
of NGC~507 in the 0.8--3~keV band is plotted for XMM-Newton MOS1+2 ($r<10'$). 
The best-fit $\beta$ model is shown by the solid gray line. 
In the bottom panel, the fit residuals are shown in units of $\sigma$, 
which correspond to the data minus the folded model, divided
by $1\sigma$ error of each data point.
}\label{fig:2}
\end{figure}

\subsection{Generation of ARFs}
\label{subsec:arf}
A precise surface brightness profile of NGC~507 was needed to generate
the Suzaku ARF, and we used the XMM-Newton image which had much better
spatial resolution than Suzaku.  For this XMM-Newton data, we used
MOS1+2 data (34.7 ks), and followed the data reduction by
\citet{sato05}.  We subtracted blank-sky data as the background
\citep{read03}, eliminated point sources in the ICM, and corrected the
vignetting.  Figure~\ref{fig:2} shows a radial profile of NGC~507 with
XMM-Newton in the 0.8--3~keV range, fitted with a $\beta$ model.  The
origin of the profile is set at (RA, Dec) = (\timeform{1h23m39.7s},
\timeform{+33D15'22''}) in J2000.0.  The best-fit parameters are
$\beta = 0.48$ and $r_c =$ \timeform{1.0'}\@.  Though the fit is not
acceptable, the ARF showed little influence on the temperature and 
abundance in the spectral fits.  We then generated two ARFs
for the spectrum of each annular sky region, $A^{\makebox{\small\sc u}}$ and
$A^{\makebox{\small\sc b}}$, which respectively assumed uniform sky
emission and $\sim 0.5^{\circ} \times 0.5^{\circ}$ size of the $\beta$-model
surface brightness profile obtained with the XMM-Newton data.  We did
not use the raw XMM-Newton image, but the smoothed image derived from
the parameters of the $\beta$-model fit to generate the ARFs, because
the raw image had a gap between the CCD chips. In this way, NGC~507
was characterized by a smooth and symmetric ICM distribution.

\begin{table*}
\caption{
Area, coverage of whole annulus, {\scriptsize SOURCE\_RATIO\_REG},
and observed/estimated counts for each annular region.
}\label{tab:3}
\centerline{
\begin{tabular}{lrrrcrrrrcrrrr}
\hline\hline
\makebox[3em][l]{Region\,$^\ast$} & \multicolumn{1}{c}{Area\makebox[0in][l]{\,$^\dagger$}} & Coverage\makebox[0in][l]{\,$^\dagger$}\hspace*{-0.5em} & \makebox[4.2em][r]{\scriptsize SOURCE\_\makebox[0in][l]{\,$^\ddagger$}}\hspace*{-0.5em} & Energy & \multicolumn{4}{c}{BI counts\makebox[0in][l]{\,$^\S$}} &$\!\!\!\!$& \multicolumn{4}{c}{FI counts\makebox[0in][l]{\,$^\S$}} \\
\cline{6-9}\cline{11-14}
& \makebox[2em][c]{(arcmin$^2$)} &      & \makebox[4.2em][r]{\scriptsize RATIO\_REG}\hspace*{-0.5em} & (keV) & OBS & NXB & CXB & $f_{\rm BGD}$ &$\!\!\!\!$& OBS & NXB & CXB & $f_{\rm BGD}$ \\
\hline\\[-2ex]
0$'$--2$'$& 12.6&100.0\%& 29.2\%& 0.4--7.1&$\!\!23,452$&$\!\!  335$&$\!\!338$&$\!\! 2.8\%$&$\!\!\!\!$&$\!\!49,445$&$\!\!509$&$\!\!794$&$\!\! 2.6\%$\\
2$'$--4$'$&37.7&100.0\%& 21.9\%& 0.4--7.1&$\!\!26,997$&$\!\!1,023$&$\!\!957$&$\!\!7.3\%$&$\!\!\!\!$&$\!\!53,622$&$\!\!1,501$&$\!\!2,284$&$\!\!7.1\%$\\
4$'$--6$'$&62.8&100.0\%& 14.4\%& 0.4--7.1&$\!\!18,274$&$\!\!1,708$&$\!\!1,405$&$\!\!17.0\%$&$\!\!\!\!$&$\!\!34,960$&$\!\!2,488$&$\!\!3,356$&$\!\!16.7\%$\\
6$'$--9$'$ &133.3& 94.3\%&  13.4\%& 0.4--7.1&$\!\!
 21,286$&$\!\!3,490$&$\!\!2,419$&$\!\!27.8\%$&$\!\!\!\!$&$\!\!38,891$&$\!\!4,961$&$\!\!5,520$&$\!\!26.9\%$\\
$r>9'$ & 64.0& 23.1\%&  4.0\%& 0.4--3.0&$\!\! 5,481$&$\!\!  927$&$\!\!  737$&$\!\!30.4\%$&$\!\!\!\!$&$\!\! 9,420$&$\!\!1,218$&$\!\!1,497$&$\!\!28.8\%$\\
\hline
\end{tabular}
}

\medskip
\parbox{\textwidth}{\footnotesize
{\scriptsize SOURCE\_RATIO\_REG} represents the
flux ratio in the assumed spatial distribution on the sky
($\beta$-model) inside the accumulation region
to the entire model, and written in the header keyword of
the calculated ARF response by ``xissimarfgen''.
}
\parbox{\textwidth}{\footnotesize
\footnotemark[$\ast$]
The outermost region ($r>9'$) include the calibration source region.

\footnotemark[$\dagger$]
The largest values among four sensors are presented.

\footnotemark[$\ddagger$]
$\makebox{\scriptsize\rm SOURCE\_RATIO\_REG}\equiv
\makebox{\scriptsize\rm COVERAGE}\;\times
\int_{r_{\rm in}}^{r_{\rm out}} S(r)\; r\,dr / 
\int_{0}^{\infty} S(r)\; r\,dr$,
where $S(r)$ represents the assumed radial profile
of NGC~507, and we defined $S(r)$ in $30'\times 30'$ region on the sky.

\footnotemark[$\S$]
OBS denotes the observed counts including NXB and CXB in 0.4--7.1~keV
or 0.4--3~keV\@. NXB and CXB are the estimated counts.
}
\end{table*}

\section{Temperature and Abundance Profiles}

\subsection{Spectral Fit}
\label{sec:spec}

\begin{figure*}
\begin{minipage}{0.33\textwidth}
\FigureFile(\textwidth,\textwidth){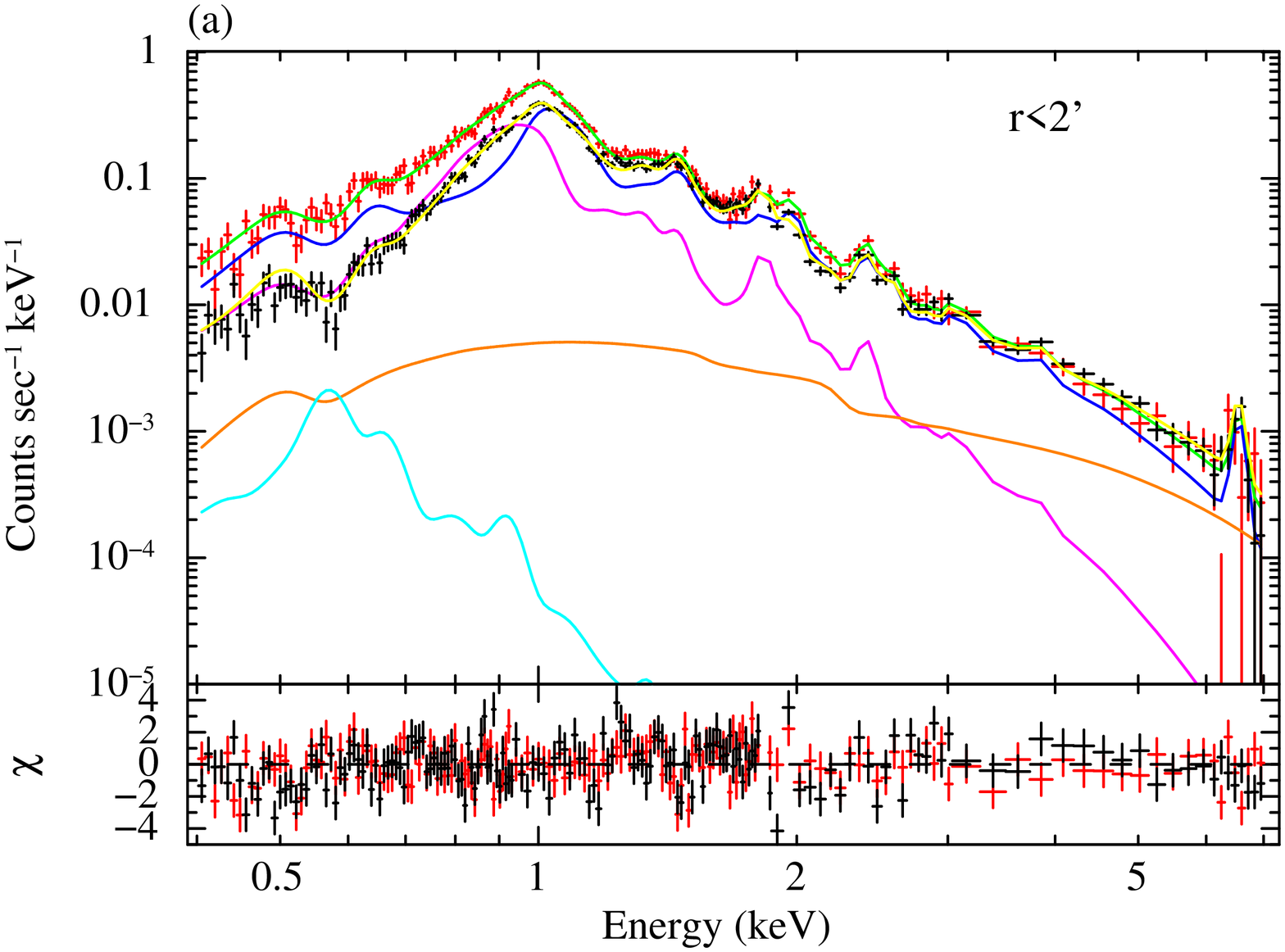}
\end{minipage}\hfill
\begin{minipage}{0.33\textwidth}
\FigureFile(\textwidth,\textwidth){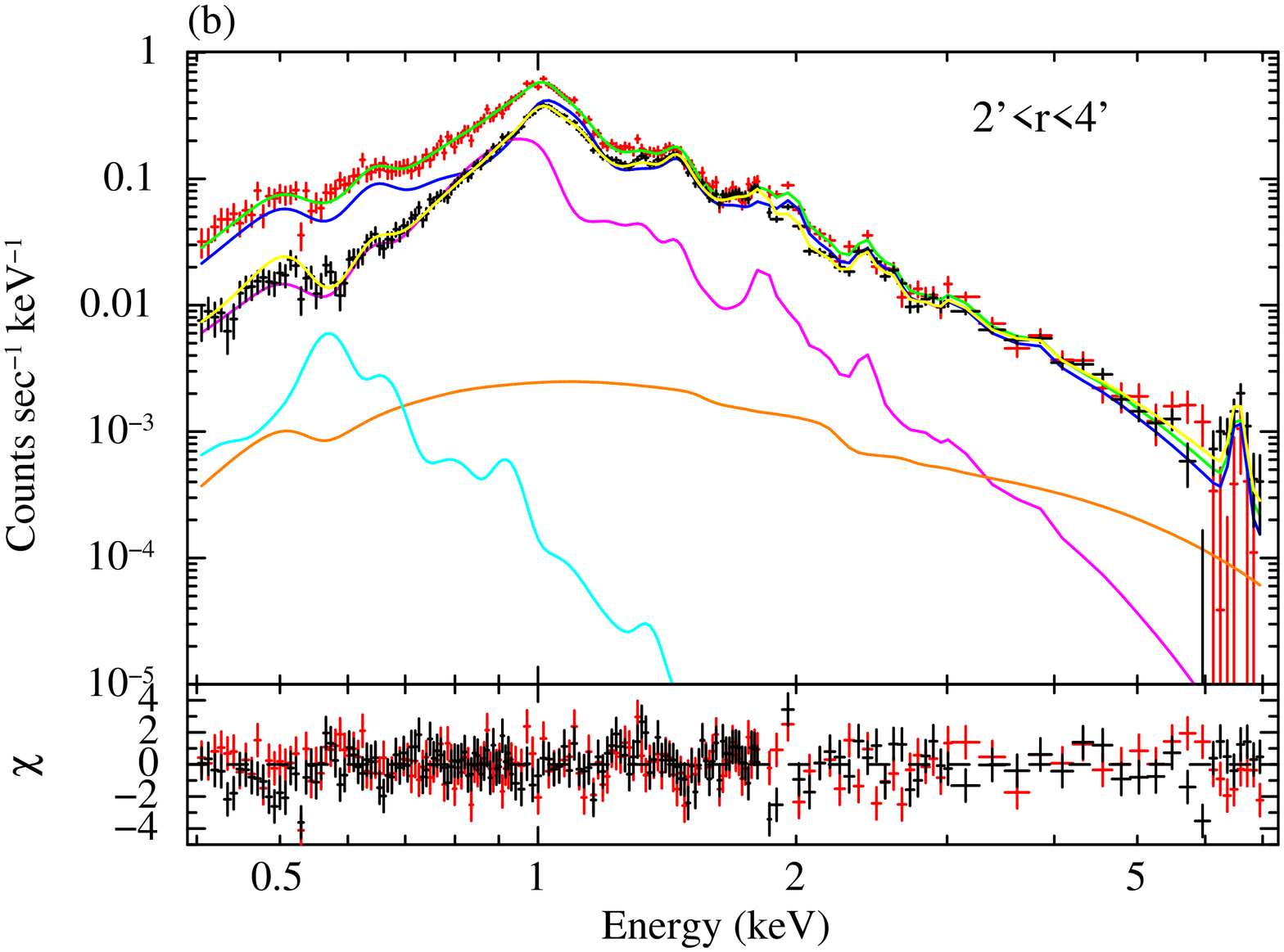}
\end{minipage}\hfill
\begin{minipage}{0.33\textwidth}
\FigureFile(\textwidth,\textwidth){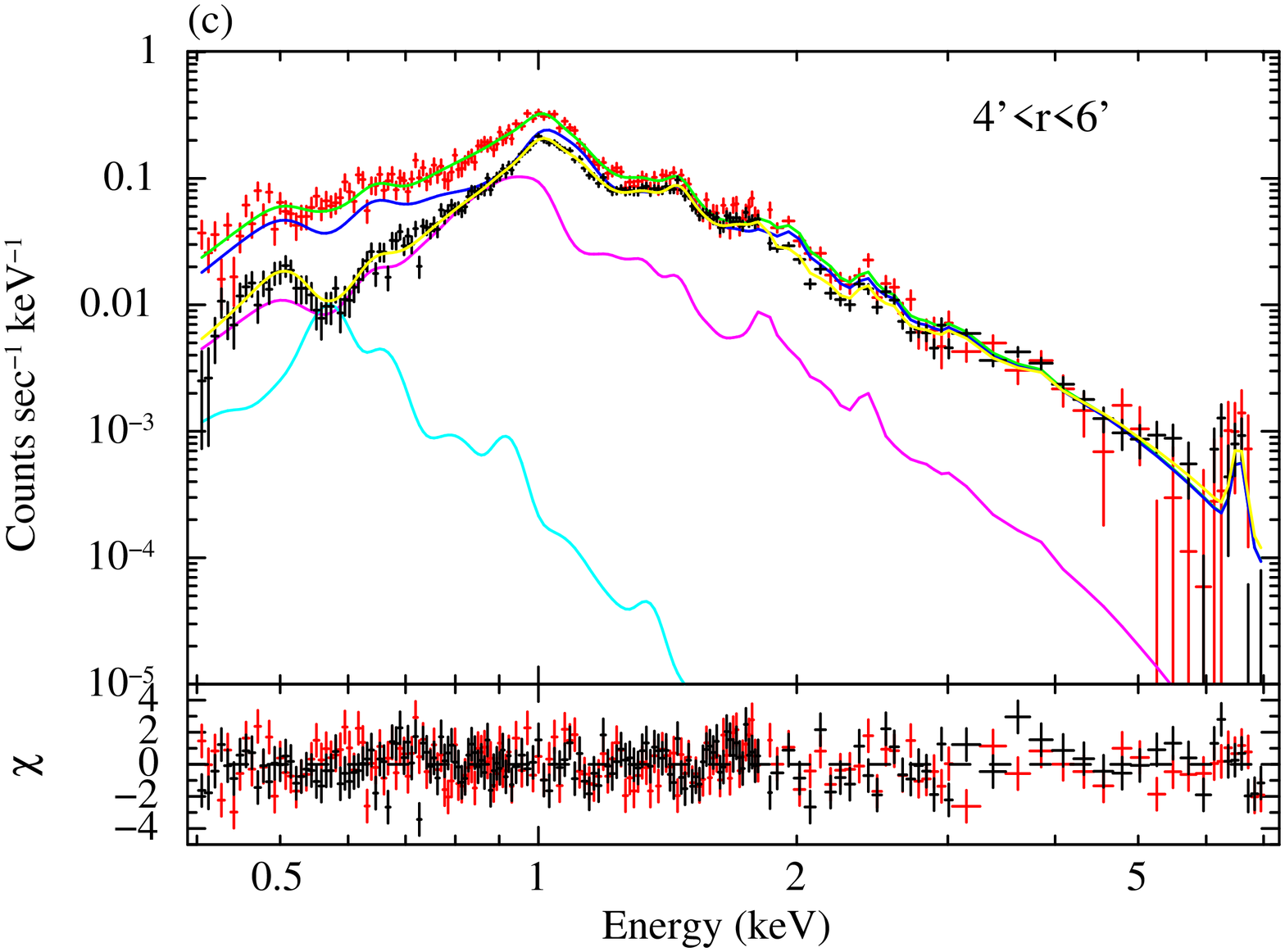}
\end{minipage}

\begin{minipage}{0.33\textwidth}
\FigureFile(\textwidth,\textwidth){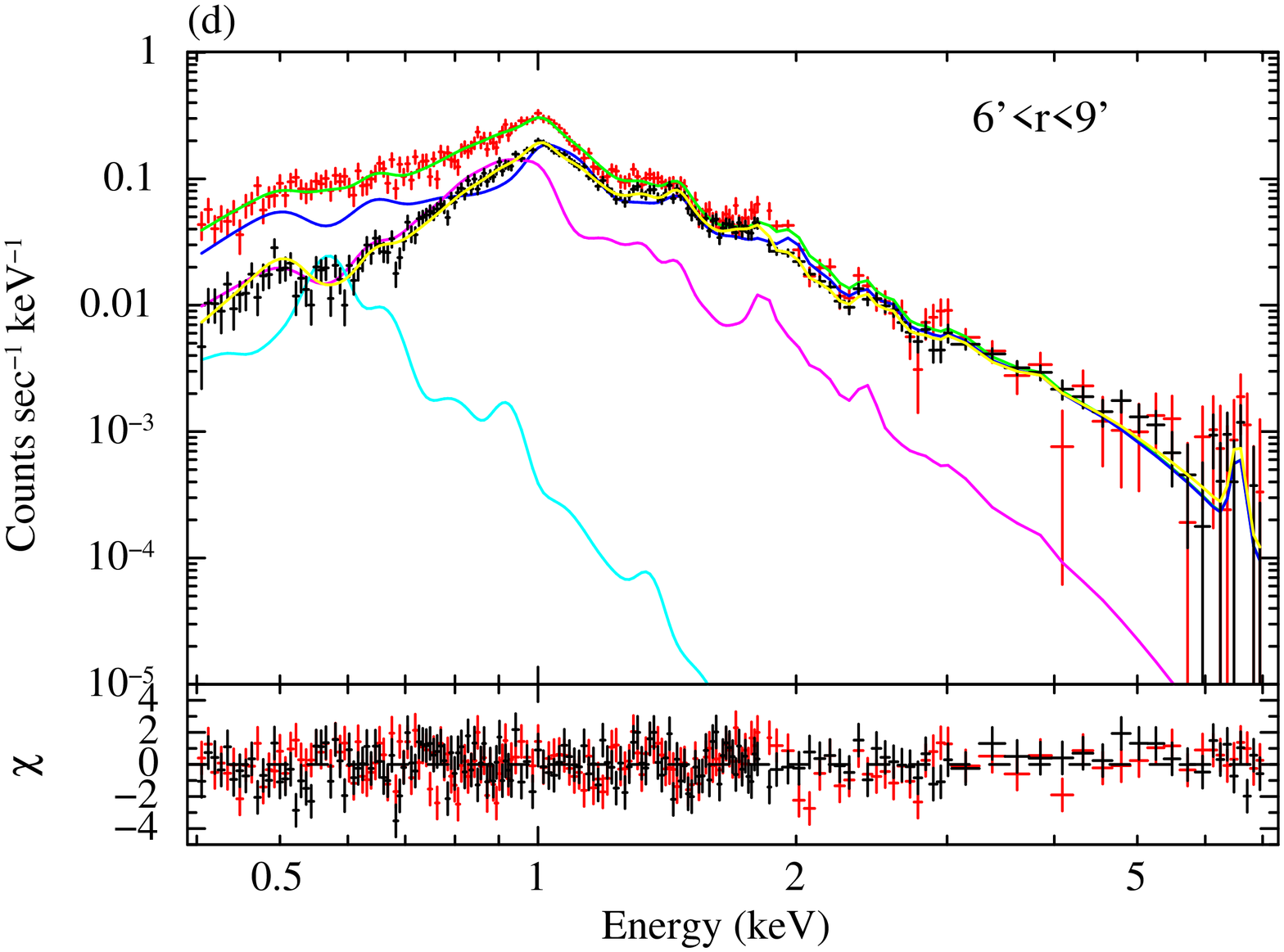}
\end{minipage}\hfill
\begin{minipage}{0.33\textwidth}
\FigureFile(\textwidth,\textwidth){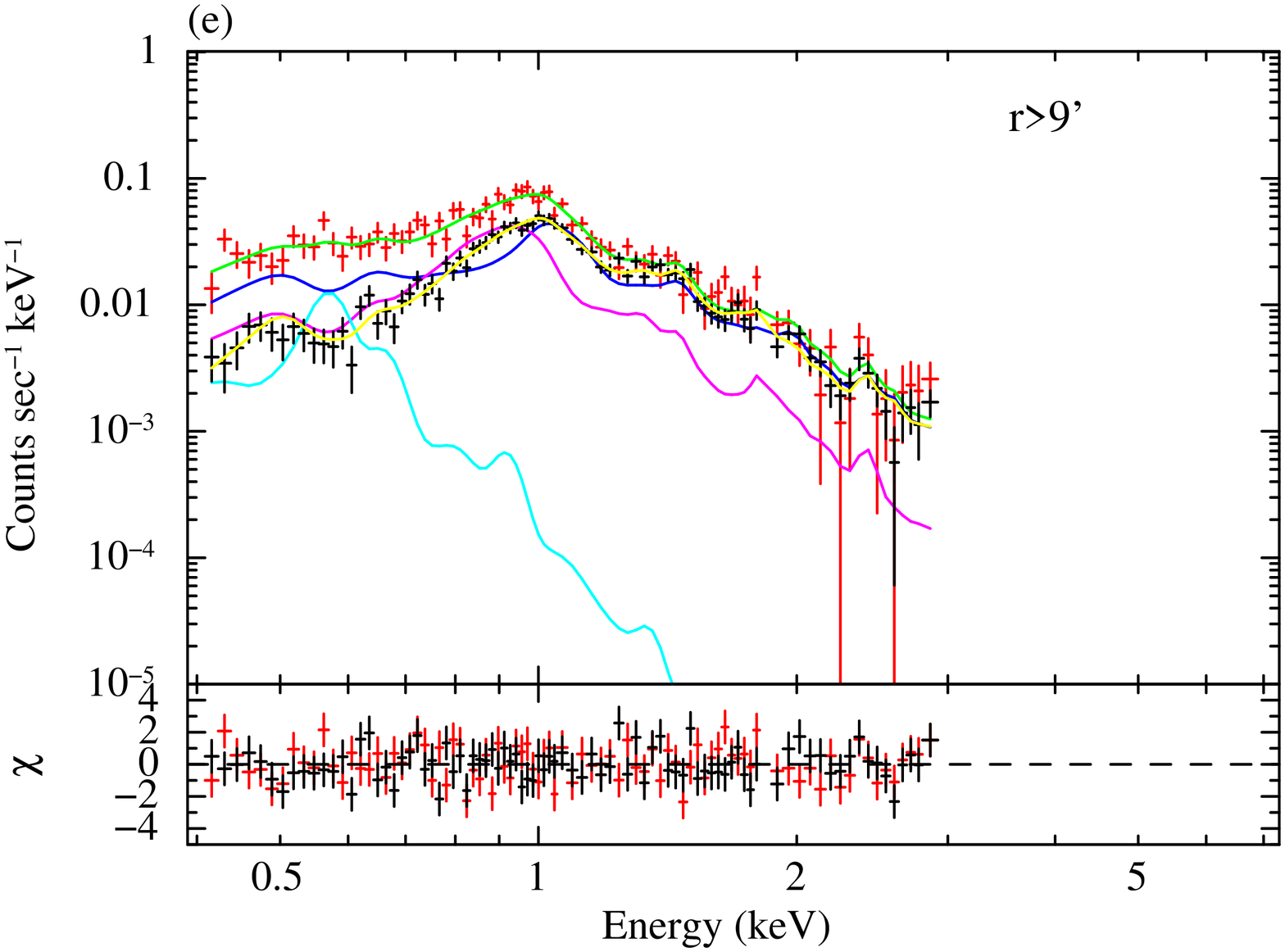}
\end{minipage}\hfill
\makebox[0.33\textwidth][l]{ }

\caption{The panels show the observed spectra
at the annular regions of NGC~507 which are denoted in the panels,
and they are plotted by red and black crosses for BI and FI, respectively.
The estimated CXB and NXB components are subtracted,
and they are fitted with the
${\it apec}+ {\it phabs} \times ({\it vapec}_1 + {\it vapec}_2 + 
{\it zbremss})$ model
shown by green and yellow lines for the BI and FI spectra.
\bf The ${\it vapec}_1$ (hot) and ${\it vapec}_2$ (cool) components 
of ICM correspond to blue and magenta lines, respectively.
The ${\it apec}$ component for the BI spectra
are indicated by cyan line.
The ${\it zbremss}$ component for the BI spectra within $r<4'$ region 
are indicated by orange line.
The energy range around the Si K-edge (1.825--1.840 keV) is ignored
in the spectral fit.
The lower panels show the fit residuals in units of $\sigma$.
}\label{fig:3}
\end{figure*}

We extracted spectra from five annular regions
of 0$'$--2$'$, 2$'$--4$'$, 4$'$--6$'$, 6$'$--9$'$, $r>9'$,
centered on (RA, Dec) = (\timeform{1h23m40.0s}, \timeform{+33D15'21''}).
Table~\ref{tab:3} lists the areas of the extracted regions (arcmin$^2$),
fractional coverage of the annulus (\%),
the {\sc source\_ratio\_reg} values (\%; see caption for its definition)
and the BI and FI counts for the observed spectra and the estimated 
NXB and CXB spectra. The fraction of the background,
$f_{\rm BGD}\equiv\rm (NXB + CXB)/OBS$, was less than $\sim$30\%
even at the outermost annulus, although the Galactic component
is not considered here.
Each annular spectrum is shown in figure~\ref{fig:3}.
The ionized Mg, Si, S, Fe lines are clearly seen in each region.
The O\emissiontype{VII} and O\emissiontype{VIII} lines are prominent
in the outer rings, however, most of the O\emissiontype{VII} line
is considered to come from the local Galactic emission,
and we dealt with those in the same way as \citet{sato07a} and 
\citet{sato08}.

\begin{table*}
\caption{
The best-fit parameters of the {\it apec} component
for the simultaneous fit of all spectra of NGC~507
with one or two temperature models (${\it apec}$) for Galactic emissions 
and ${\it phabs} \times ({\it vapec}_{\rm 1T~or~2T} + 
 {\it zbremss}_{r<4'})$ model for ICM.
}\label{tab:4}
\begin{tabular}{lccccc}
\hline\hline
\makebox[19em][l]{Fit model} & ${\it Norm}_1\,^\ast$ & $kT_1$ & ${\it Norm}_2\,^\ast$ & $kT_2$ &  $\chi^2$/dof\\
 & & (keV) &\\
\hline
 ${\it apec}_1 + {\it phabs}\times ({\it vapec}_1 + {\it zbremss})$ & 0.47 & 0.110 &-- & -- & 3173/1520\\
 ${\it apec}_1 + {\it phabs}\times ({\it vapec}_1 + {\it vapec}_2 + {\it zbremss})$ & $0.42\pm 0.14$ & $0.159^{+0.034}_{-0.037}$ &-- & -- & 2067/1510\\
 ${\it apec}_1 +{\it apec}_2 + {\it phabs}\times ({\it vapec}_1 + {\it zbremss})$ & 0.26 & 0.1 (fix) & 0.33  & 0.3 (fix) & 3073/1520\\
 ${\it apec}_1 +{\it apec}_2 + {\it phabs}\times ({\it vapec}_1 + {\it vapec}_2 + {\it zbremss})$ & $1.02\pm 0.42$ & 0.1 (fix) & $0.13\pm 0.13$  & 0.3 (fix) & 2075/1510\\
\hline\\[-1ex]
\multicolumn{6}{l}{\parbox{0.94\textwidth}{\footnotesize 
\footnotemark[$*$] 
Normalization of the {\it apec} component
divided by the solid angle, $\Omega^{\makebox{\tiny\sc u}}$,
assumed in the uniform-sky ARF calculation (20$'$ radius),
${\it Norm} = \int n_{\rm e} n_{\rm H} dV \,/\,
(4\pi\, (1+z)^2 D_{\rm A}^{\,2}) \,/\, \Omega^{\makebox{\tiny\sc u}}$
$\times 10^{-20}$ cm$^{-5}$~arcmin$^{-2}$, 
where $D_{\rm A}$ is the angular distance to the source.}}
\end{tabular}
\end{table*}

\begin{figure}
\begin{center}
\begin{minipage}{0.45\textwidth}
\FigureFile(\textwidth,\textwidth){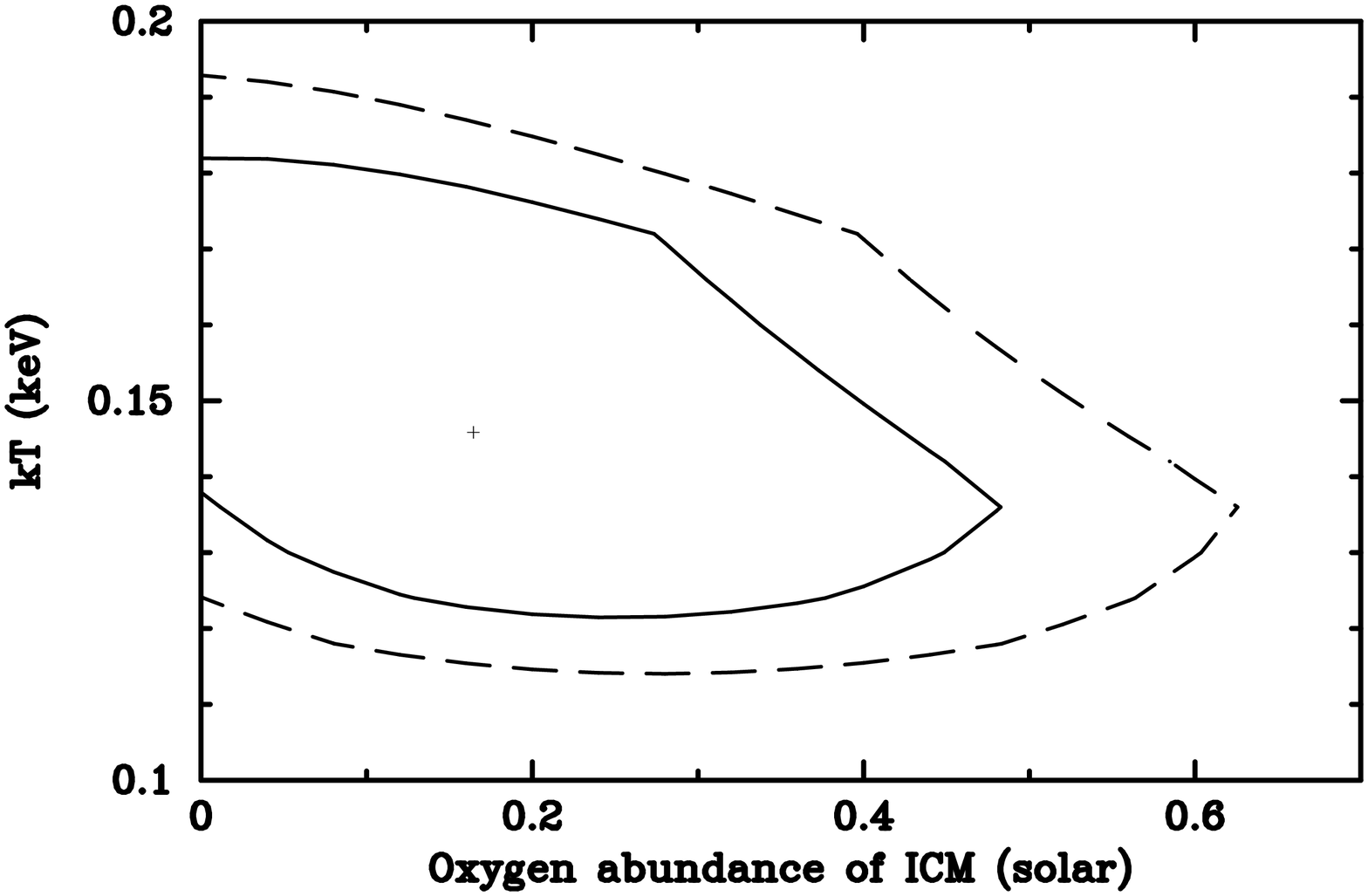}
\caption{A plot of confidence contour between $kT$ of {\it apec} component
(i.e. Galactic component) and the O abundance of {\it vapec} model 
(i.e. ICM) for the $r>9'$ annulus,
in the simultaneous fitting of all annuli with the
${\it apec} + {\it phabs} \times ({\it vapec}_1 + {\it vapec}_2 
+ {\it zbremss}) $ model.
The cross denotes the best-fit location,
and the two contours represent 1$\sigma$ and 90\% confidence
ranges, from inner to outer, respectively.
}\label{fig:4}
\end{minipage}
\end{center}
\end{figure}

\begin{figure*}
\begin{minipage}{0.33\textwidth}
\FigureFile(\textwidth,\textwidth){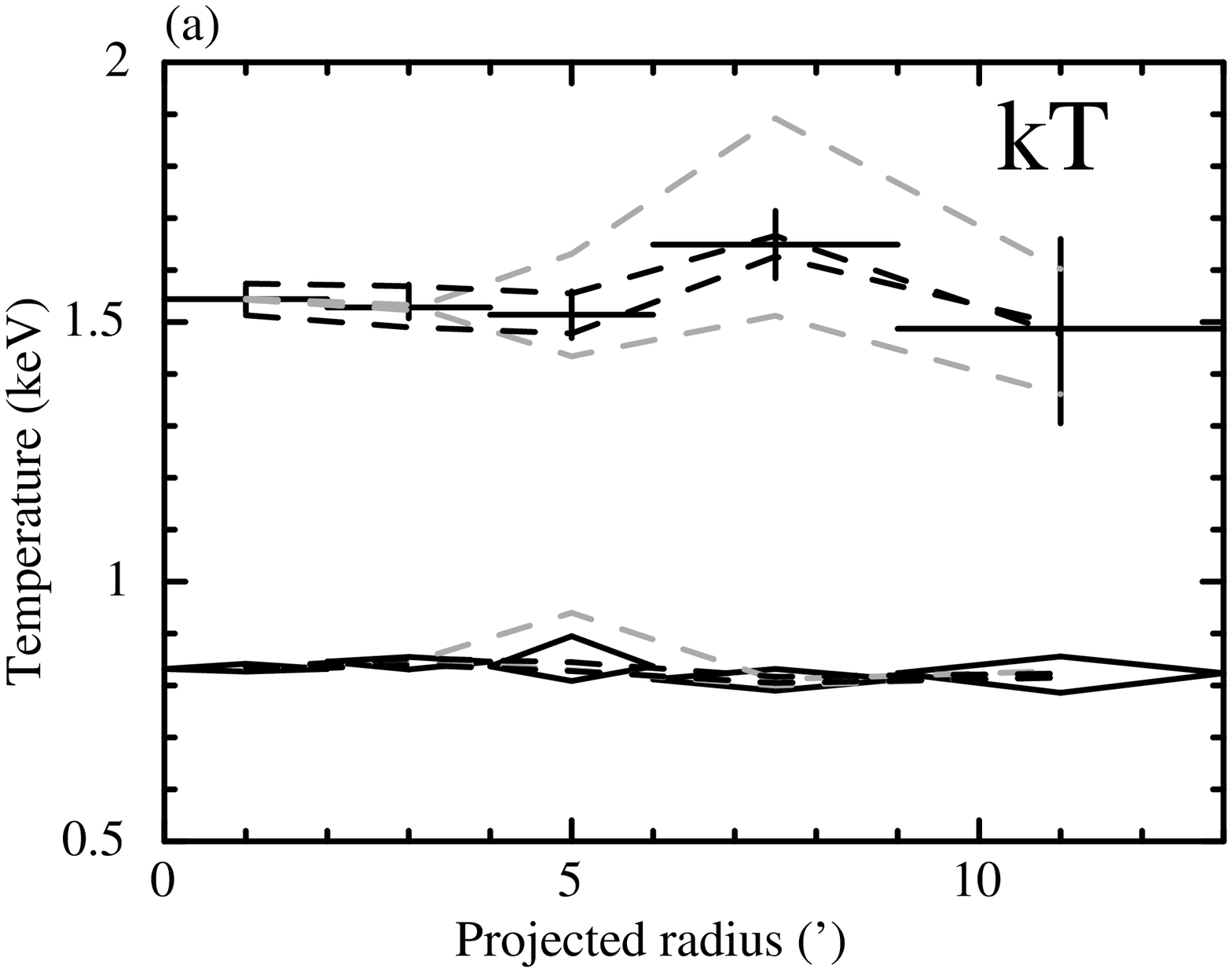}
\end{minipage}\hfill
\begin{minipage}{0.33\textwidth}
\FigureFile(\textwidth,\textwidth){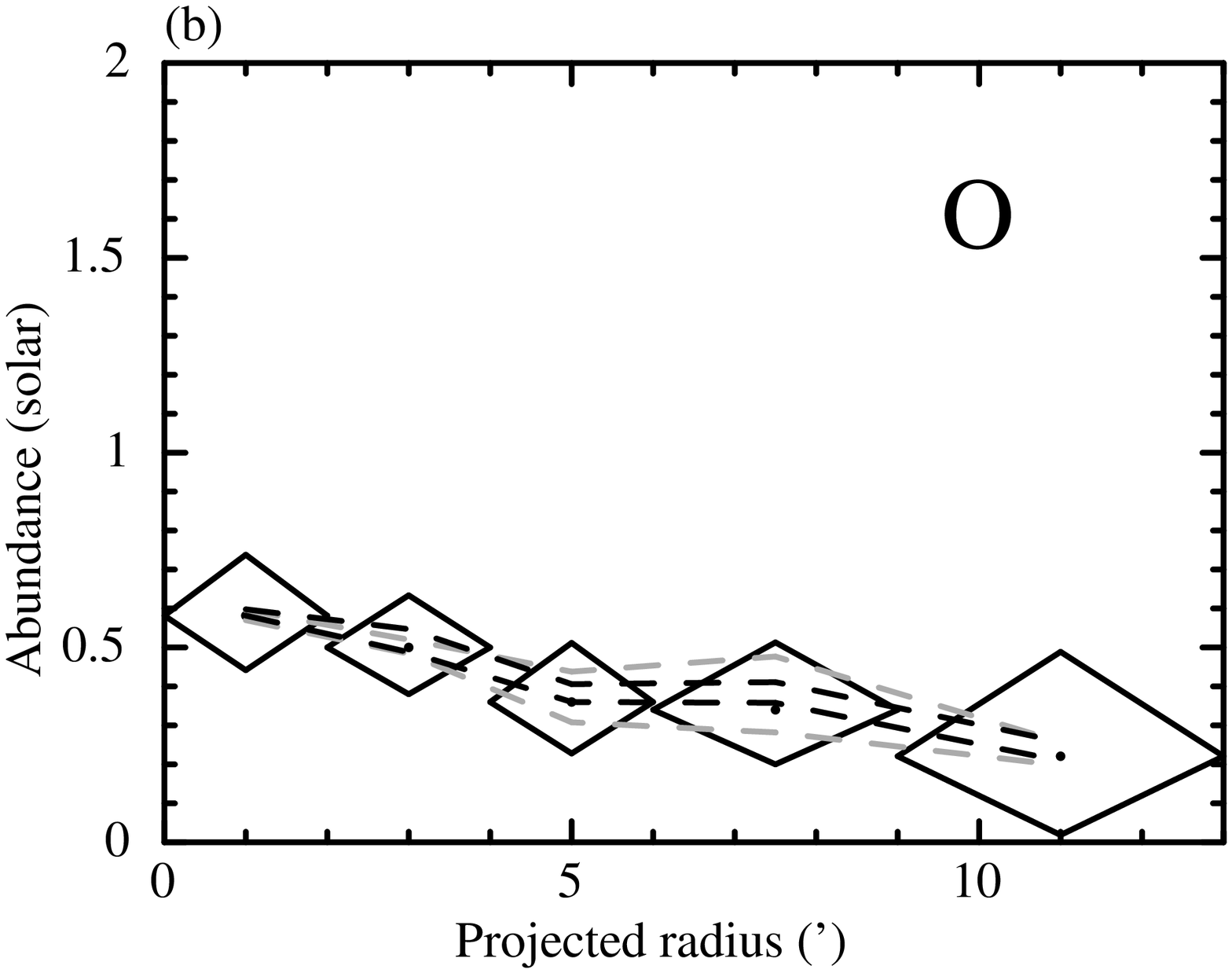}
\end{minipage}\hfill
\begin{minipage}{0.33\textwidth}
\FigureFile(\textwidth,\textwidth){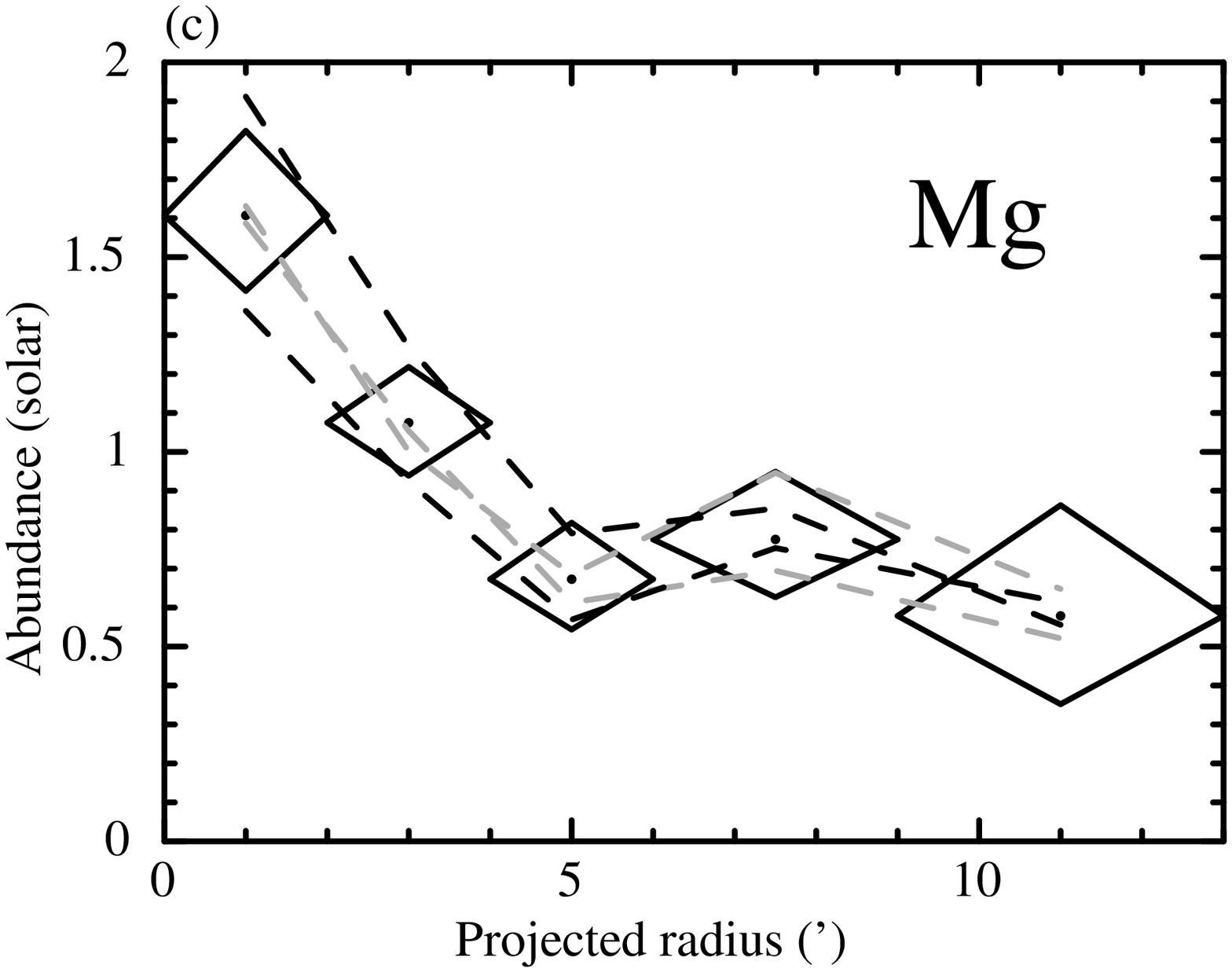}
\end{minipage}

\begin{minipage}{0.33\textwidth}
\FigureFile(\textwidth,\textwidth){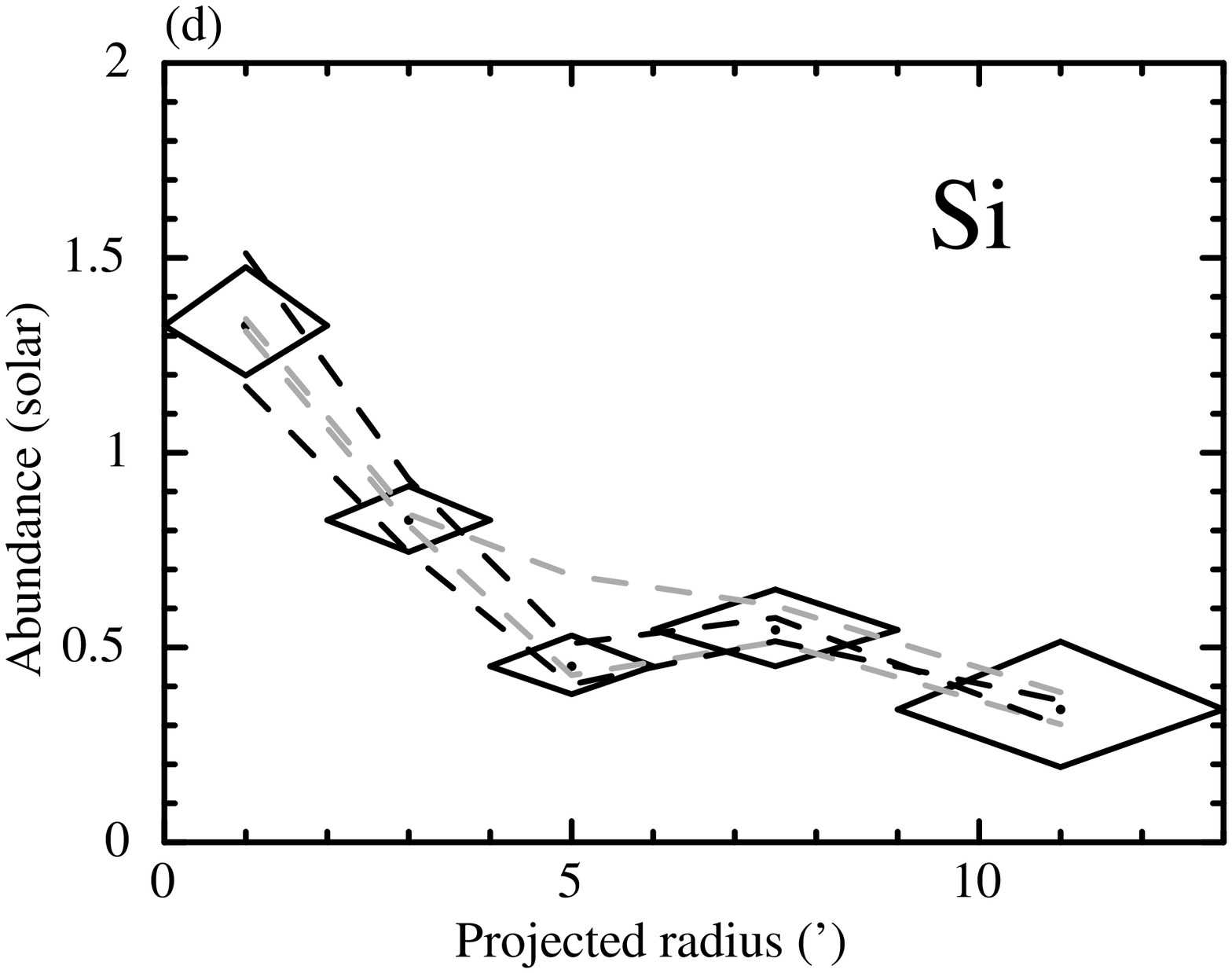}
\end{minipage}\hfill
\begin{minipage}{0.33\textwidth}
\FigureFile(\textwidth,\textwidth){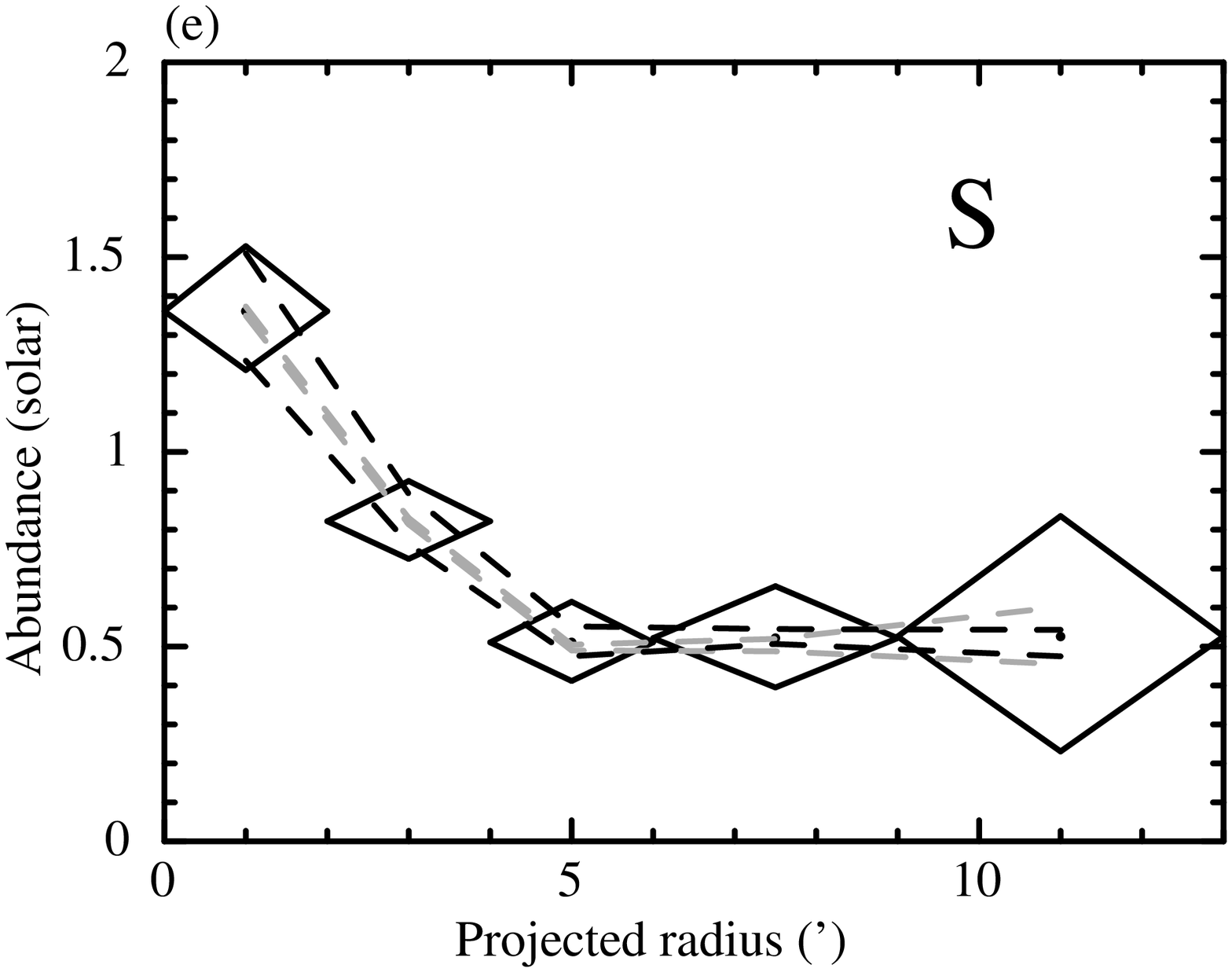}
\end{minipage}\hfill
\begin{minipage}{0.33\textwidth}
\FigureFile(\textwidth,\textwidth){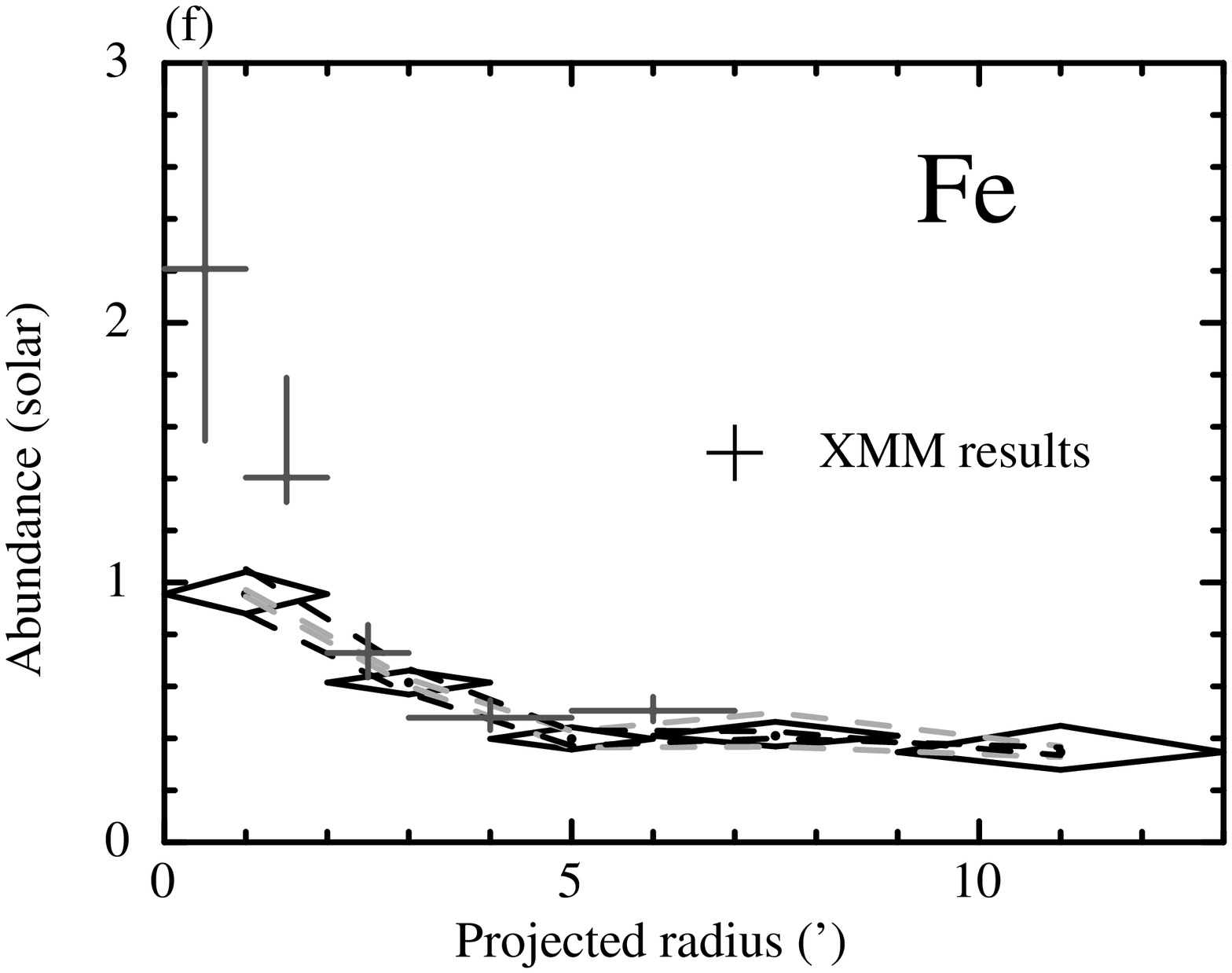}
\end{minipage}

\vspace*{-0.5ex}
\caption{(a): Radial temperature profiles derived from the spectral fit
for each annulus against the projected radius.
Black dashed lines show systematic change of the best-fit values
by varying the thickness of the OBF contaminant by $\pm 10$\%.
Light-gray dashed lines denote those when the estimated CXB and NXB
levels are varied by $\pm 10$\%.  
(b)--(f): Radial abundance profiles derived and plotted in the same way 
as in (a).  (f): XMM results of Fit 13 in \citet{kim04} correspond to 
gray crosses.}
\label{fig:5}
\end{figure*}

\begin{table*}
\caption{Summary of the parameters from one or two {\it vapec} fit
(${\it apec} + {\it phabs} \times ({\it vapec}_{\rm 1T~or~2T} + 
{\it zbremss}_{r<4'})$) to
each annular spectrum of NGC~507.  All annuli were simultaneously
fitted.  Errors are 90\% confidence range of statistical errors, and
do not include systematic errors.  The solar abundance ratio of {\it
angr} was assumed.  These results are plotted in figure~\ref{fig:5}.
}\label{tab:5}
\begin{center}
\begin{tabular}{lcccccc}
\hline\hline
\makebox[2em][l]{ICM1T} & {\it Norm}$_1$\makebox[0in][l]{\,$^\ast$} & $kT_1$& {\it Norm}$_2$\makebox[0in][l]{\,$^\ast$} & $kT_2$& {\it Norm}$_1$/{\it Norm}$_2$ & $\chi^2$/dof \\
& &(keV)&&(keV)& & \\
\hline
0--2$'$    & 527.0 & 1.21 & -- & -- & -- &  1118/339\\
2--4$'$    & 99.6  & 1.27 & -- & -- & -- &  620/339\\
4--6$'$    & 29.0 & 1.26 & -- & -- & -- &  560/340\\
6--9$'$    & 9.3 & 1.17 & -- & -- & -- &  638/340\\
$r>9'$     & 3.8 & 1.11 & -- & -- & -- &  237/162\\
total      &  &  &    &    &    & 3173/1520\\
\hline
\makebox[2em][l]{ICM1T} &O&Ne&Mg,Al&Si&\makebox[0in][c]{S,Ar,Ca}&Fe,Ni \\
&(solar)&(solar)&(solar)&(solar)&(solar)&(solar) \\
\hline
0--2$'$    & 0.23 & 0.00 & 0.25 & 0.31 & 0.40 & 0.22 \\
2--4$'$    & 0.26 & 0.23 & 0.33 & 0.31 & 0.43 & 0.26 \\
4--6$'$    & 0.35 & 0.05 & 0.56 & 0.55 & 0.63 & 0.42 \\
6--9$'$    & 0.36 & 0.00 & 0.59 & 0.68 & 0.80 & 0.51 \\
$r>9'$     & 0.21 & 0.00 & 0.22 & 0.19 & 0.41 & 0.18 \\
\hline\hline
\makebox[2em][l]{ICM2T} & {\it Norm}$_1$\makebox[0in][l]{\,$^\ast$} & $kT_1$& {\it Norm}$_2$\makebox[0in][l]{\,$^\ast$} & $kT_2$& {\it Norm}$_1$/{\it Norm}$_2$ & $\chi^2$/dof \\
& &(keV)&&(keV)& & \\
\hline
0--2$'$    &  $115.4\pm7.0$  & 1.54$^{+0.02}_{-0.02}$ &  $38.0\pm3.2$  & 0.83$^{+0.01}_{-0.01}$ & $3.04\pm0.32$ &  576/337\\
2--4$'$    &  $49.0\pm2.8$  & 1.53$^{+0.04}_{-0.02}$ &  $11.3\pm0.8$  & 0.85$^{+0.01}_{-0.01}$ & $4.34\pm0.39$ & 436/337\\
4--6$'$    &  $26.1\pm1.7$  & 1.51$^{+0.05}_{-0.04}$ &  $5.7\pm0.6$  & 0.84$^{+0.06}_{-0.03}$ & $4.58\pm0.57$ & 469/338\\
6--9$'$    &  $12.6\pm0.9$  & 1.65$^{+0.06}_{-0.06}$ &  $4.2\pm0.4$  & 0.81$^{+0.02}_{-0.02}$& $3.00\pm0.36$ & 399/338\\
$r>9'$    &  $8.5\pm1.2$  & 1.49$^{+0.17}_{-0.18}$ &  $4.1\pm0.8$  & 0.82$^{+0.03}_{-0.04}$ & $2.07\pm0.54$ & 187/160\\
total  &  &  &  &  &  & 2067/1510\\\\[-2ex]
\hline
\makebox[2em][l]{ICM2T} &O&Ne&Mg,Al&Si&\makebox[0in][c]{S,Ar,Ca}&Fe,Ni \\
&(solar)&(solar)&(solar)&(solar)&(solar)&(solar) \\
\hline
0--2$'$    & 0.58$^{+0.16}_{-0.14}$ & 2.02$^{+0.41}_{-0.36}$ & 1.61$^{+0.22}_{-0.19}$ & 1.33$^{+0.15}_{-0.13}$ & 1.36$^{+0.17}_{-0.15}$ & 0.96$^{+0.09}_{-0.08}$ \\
2--4$'$    & 0.50$^{+0.13}_{-0.12}$ & 1.34$^{+0.20}_{-0.20}$ & 1.08$^{+0.14}_{-0.14}$ & 0.83$^{+0.09}_{-0.08}$ & 0.82$^{+0.10}_{-0.10}$ & 0.61$^{+0.05}_{-0.05}$ \\
4--6$'$    & 0.36$^{+0.15}_{-0.13}$ & 1.00$^{+0.28}_{-0.26}$ & 0.67$^{+0.15}_{-0.13}$ & 0.45$^{+0.08}_{-0.07}$ & 0.51$^{+0.10}_{-0.10}$ & 0.40$^{+0.05}_{-0.04}$ \\
6--9$'$    & 0.34$^{+0.17}_{-0.14}$ & 1.19$^{+0.32}_{-0.29}$ & 0.77$^{+0.17}_{-0.15}$ & 0.54$^{+0.10}_{-0.10}$  & 0.52$^{+0.13}_{-0.13}$ & 0.41$^{+0.05}_{-0.04}$ \\
$r>9'$    & 0.22$^{+0.27}_{-0.20}$ & 0.39$^{+0.52}_{-0.39}$ & 0.58$^{+0.28}_{-0.23}$ & 0.34$^{+0.17}_{-0.15}$  & 0.53$^{+0.31}_{-0.29}$ & 0.35$^{+0.10}_{-0.07}$  \\
\hline\\[-1ex]
\multicolumn{7}{l}{\parbox{0.75\textwidth}{\footnotesize
\footnotemark[$\ast$] 
Normalization of the {\it vapec} component scaled with a factor of
{\sc source\_ratio\_reg} / {\sc area} in table~\ref{tab:3},\\ ${\it
Norm}=\frac{\makebox{\sc source\_ratio\_reg}}{\makebox{\sc area}} \int
n_{\rm e} n_{\rm H} dV \,/\, [4\pi\, (1+z)^2 D_{\rm A}^{\,2}]$ $\times
10^{-20}$~cm$^{-5}$~arcmin$^{-2}$, where $D_{\rm A}$ is the angular
distance to the source. }}\\
\multicolumn{7}{l}{\parbox{0.75\textwidth}{\footnotesize
\footnotemark[$\dagger$] 
All regions were fitted simultaneously.
}}
\end{tabular}
\end{center}
\end{table*}

The spectra from BI and FI for all regions were fitted simultaneously.
For the four inner regions the fitted energy range was 0.4--7.1 keV,
and it was 0.4--3.0 keV for the outermost region. 
In the simultaneous fit, the Galactic emission component as a background 
was common for all regions, while the NGC~507 emission component 
was unlinked between each region. 
We excluded the narrow energy band around the Si K-edge (1.825--1.840~keV) 
because of incomplete response.  The energy range below 0.4~keV was also
excluded because the C edge (0.284~keV) seen in the BI spectra could
not be reproduced well in our data.  The range above 7.1~keV was also
ignored because Ni line ($\sim 7.5$~keV) in the background left an
artificial structure after the NXB subtraction at large radii. 
In the simultaneous fit of BI and FI data, only normalization parameters were
allowed to be different between them, although we found that the
derived normalizations were quite consistent to be the same.  For the
outermost region ($r>9'$) we fitted the spectrum in 0.4--3.0 keV to
exclude contribution from the Fe-55 calibration source.

It is important to estimate the Galactic component precisely. 
In order to determine surface brightness and
spectral shape of the Galactic component, we carried out simultaneous
fit for all annuli.  The Galactic component gives significant
contribution in these annuli as shown in figure~\ref{fig:3}, however
the ICM component is still dominant in almost all the energy range
except for the O\emissiontype{VII} line.

We assumed either one or two temperature {\it apec} model for the
Galactic component, and fitted the data with the following model
formula: ${\it apec}_{\rm 1T~or~2T} + {\it phabs}\times ({\it
vapec}_{\rm 1T~or~2T} + {\it zbremss})$. 
The resultant normalizations of the {\it apec} models 
in table~\ref{tab:4} are scaled so that they give the surface brightness 
in the unit solid angle of arcmin$^2$, and the normalizations of 
the ${\it apec}$ components were constrained to give 
the same surface brightness and the same temperature for all annuli.
When we fitted the data with two {\it apec} models whose temperatures 
were free parameters, we could not obtain reasonable fit. Thus, 
we fixed the two temperatures to be 0.1 and 0.3 keV, which were 
consistent with the previous studies (e.g.\ \cite{lumb02}).  
When we assumed two {\it apec} model, whose
temperatures were fixed to be 0.1 and 0.3 keV, the resultant
parameters were almost the same as those in table \ref{tab:5}.
As a result, we concluded that only one {\it apec} model was enough to 
fit the NGC~507 data since the fit
improvement showed low significance with the two {\it apec} model. 
Therefore, we performed simultaneous fit in the 0.4--7.1~keV range 
(excluding 1.825--1.840~keV) for the four inner regions and 
in the 0.4--3.0~keV range for the $r>9'$ region, assuming 
an {\it apec} model for the Galactic component, 
by following model, ${\it apec} + {\it phabs} \times ({\it vapec}_1 
+ {\it vapec}_2 + {\it zbremss}) $, as shown in table~\ref{tab:4}.  

To demonstrate how sensitive the O abundance of the ICM is to the
assumed Galactic component model, we show in figure~\ref{fig:4} a
confidence contour between $kT$ (keV) of the ${\it apec}$ component
and the O abundance (solar) of ${\it vapec}$ for the outermost annulus
($r>9'$).  There seems a negative correlation between the two
parameters, because higher temperature of the Galactic component
produces more intense O\emissiontype{VIII} line relative to
O\emissiontype{VII} one, which pushes to reduce the
O\emissiontype{VIII} line from the ICM ({\it vapec} component).
Influence on the derived temperature and abundance by the modeling of
the Galactic component will be examined in subsection
\ref{subsec:radial}, too.

The ICM spectra for all regions were clearly better represented by two
{\it vapec} models than one {\it vapec} model as shown in
table~\ref{tab:5}.  The abundances were linked in the following way,
Mg=Al, S=Ar=Ca, Fe=Ni, and were also linked between the two ${\it
vapec}$ components for each region.  
The fit results are shown in table \ref{tab:5}.
\citet{kim04} showed that the model for the central region needed to
include the low-mass X-ray binary (LMXB) component based on the ROSAT
and XMM-Newton observations, we included a $kT=7$ keV {\it zbremss}
model for this component in the fit for 0--2$'$ and 2--4$'$ annuli.
The resultant flux in 0.3--8 keV was $\sim1\times10^{-13}$ and $\sim
4\times10^{-14}$ erg s$^{-1}$ cm$^{-2}$ for 0--2$'$ and 2--4$'$
region, respectively, and the flux due to the LMXB component with
Suzaku was consistent with the \citet{kim04} result.

\begin{figure}
\begin{center}
\begin{minipage}{0.45\textwidth}
\centerline{
\FigureFile(\textwidth,\textwidth){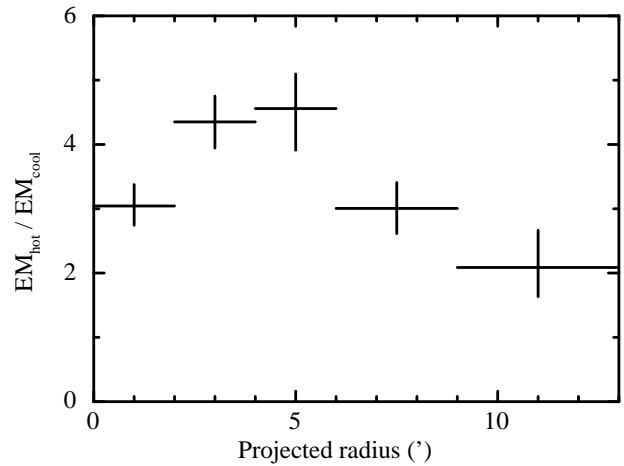}}
\caption{ The ratio of emission measure of the hot ICM component to that
of the cool ICM component.  }\label{fig:6}
\end{minipage}
\end{center}
\end{figure}

Although the fit was not statistically acceptable due mainly to the
very high photon statistics compared with the systematic errors in the
instrumental response, the results were useful to assess whether each
element abundance was reasonably determined or not.  Abundance of Ne
was not reliably determined due to the strong and complex Fe-L line
emissions, however we left these abundance to vary freely during the
spectral fit.

Results of the spectral fit for each annulus are summarized in
table~\ref{tab:6} and figure~\ref{fig:4}, in which systematic error
due to the OBF contamination and background (CXB + NXB) estimation are
included.  We examined the results by changing the background
normalization by $\pm 10$\%, and the error range is plotted with
light-gray dashed lines in figure~\ref{fig:4}. The systematic error
due to the background estimation is almost negligible.  The other
systematic error concerning the uncertainty in the OBF contaminant is
shown by black dashed lines.  A list of $\chi^2$/dof is presented in
table \ref{tab:6}.

\begin{table*}
\caption{List of $\chi^2$/dof for each fit of NGC~507.}
\label{tab:6}
\begin{center}
\begin{tabular}{lccccc}
\hline\hline
\makebox[6em][l]{Region} & nominal &\multicolumn{2}{c}{contaminaut} & \multicolumn{2}{c}{background}\\
\hline
 & & \makebox[0in][c]{+10\%} & \makebox[0in][c]{-10\%} & \makebox[0in][c]{+10\%} & \makebox[0in][c]{-10\%}\\
\hline
All $\dotfill$   & 2067/1510 & 2004/1510 & 2148/1510 & 2104/1510 & 2094/1510 \\
\hline
\end{tabular}
\end{center}
\end{table*}

\subsection{Temperature Profile}
\label{subsec:radial}

As mentioned in the previous section, the spectral fits needed 
two temperature model for ICM rather than one temperature model.
As shown in table \ref{tab:5}, the metal abundances changed 
dramatically whether we use one or two temperature models at the 
central region. 
Radial temperature profile and the ratio of the {\it vapec}
normalizations between the hot and cool ICM components are shown in
figure~\ref{fig:5}(a) and figure~\ref{fig:6}, respectively.  Our
results for the two temperature ICM model are consistent with the
XMM-Newton results \citep{kim04} for the central region within $4'$.
In the Chandra case, \citet{humphrey06} fitted the spectra with 1T
model.  The previous XMM-Newton and Chandra results neglected the
Galactic component, while our analysis took it into account.  We can
crudely approximate the two temperatures as $kT_{\rm Hot}\sim
1.5$~keV, and $kT_{\rm Cool}\sim 0.8$~keV, respectively.  The cool
component is strongest in the innermost region, and even though it
seems to decline in the outer regions, the possible coupling with the
Galactic emission makes the precise estimation difficult.  The radius
of $10'\sim 200$~kpc corresponds to $\sim 0.19\; r_{\rm 180}$, and the
temperature decline, observed in several other clusters, is not
clearly recognized in this system due partly to the multi-phase nature
of ICM\@. The temperatures of cool and hot components are consistent 
with the results of two temperature model for NGC~5044 group 
with XMM-Newton in \citet{buote03a}.

Systematic error caused by the background subtraction and by the
estimation of the XIS filter contamination was estimated.  We varied
the sum of the NXB and CXB flux and the contamination thickness
individually by $\pm 10\%$.  The results are shown by dashed lines in
figure~\ref{fig:5}(a).  Though $kT_{\rm cool}$ component is not
affected by these uncertainties, $kT_{\rm Hot}$ shows a significant
dependence on the background uncertainty most notably in the
6$'$--9$'$ annulus.  The tendency looks quite similar to the result for
HCG~62 with Suzaku \citep{tokoi08}.

\subsection{Abundance Profiles}

Metal abundances are determined for the six element groups
individually as shown in figures~\ref{fig:5}(b)--(f).  O abundance is
strongly affected by the Galactic emission as shown
figure~\ref{fig:4}.  However, the use of one or two temperature model
for the Galactic emission did not significantly affect the abundance
results as mentioned in subsection 3.1\@.  We also note that Ne
abundance has a problem in the spectral fit due to the coupling with
Fe-L lines. Therefore, regarding the spatial structure, we dealt with
the remaining four elemental groups: Mg, Si, S and Fe.  The four
abundance values and their variation with radius look quite similar to
each other. The central abundances lie between 1.0--1.6 solar, and the
abundance decline to about 1/5 of the central value is commonly seen
in the $r>9'$ annulus.

\citet{kim04} analyzed XMM-Newton data and employed the solar
abundance by \citet{grevesse98} giving $\rm [Fe/H] = 3.16\times
10^{-5}$. This was scaled here by a factor of 0.7 to match the
\citet{anders89} value of $\rm [Fe/H] = 4.68\times 10^{-5}$.  As shown
in figure~\ref{fig:5} (f), our result shows lower Fe abundance than
the \citet{kim04} value within $r<2'$, while both results are almost
consistent in $2'<r<5'$ region.  The spectral model is the same for
both analysis, employing the two temperature ICM model with the LMXB
component. The difference is that our result was derived for the
projected spectrum while \citet{kim04} used the deprojected spectra.
The Fe abundance with Suzaku is also consistent with the Chandra
result by \citet{humphrey06}. The resultant abundance profiles with 
Suzaku for NGC~507 are also consistent with those with the two 
temperature model for NGC~5044 with XMM-Newton in \citet{buote03b}.

Again, we looked into the effect of the error by the NXB and CXB
intensities and the OBF contamination. As shown by dashed lines in
figures~\ref{fig:5}(b)--(f), the systematic effect is less than the
statistical error for all regions.

\subsection{Direct Comparison of O\emissiontype{VII} and O\emissiontype{VIII} Intensities}
\label{subsec:direct}

\begin{table*}
\caption{ Intensities of O\emissiontype{VII} and O\emissiontype{VIII}
lines for each annulus of NGC~507 field in units of
photons~cm$^{-2}$~s$^{-1}$~sr$^{-1}$.  These intensities are derived
from the spectral fit with the model {\it power-law} + {\it gaussian} + {\it
gaussian}, assuming the uniform-sky ARF response.  Results
by \citet{mccammon02} measured with a high resolution microcalorimeter
array for a large sky area of $\sim 1$~sr are also presented.
}\label{tab:7}
\begin{center}
\begin{tabular}{llr} 
\hline \hline
\makebox[14em][l]{Region} & \multicolumn{1}{c}{O\emissiontype{VII}} & \multicolumn{1}{c}{O\emissiontype{VIII}} \\
\hline
\makebox[6em][l]{NGC~507} (0--2$'$)   $\dotfill$ & 14.5$^{+82.8}_{-14.5}$  & 26.0$\pm$8.4 \\
\makebox[6em][l]{NGC~507} (2--4$'$)   $\dotfill$ & 8.9$\pm$4.0  & 11.3$\pm$4.0 \\
\makebox[6em][l]{NGC~507} (4--6$'$)   $\dotfill$ & 2.5$\pm$2.5  & 3.1$\pm$2.4 \\
\makebox[6em][l]{NGC~507} (6--9$'$)   $\dotfill$ & 4.4$\pm$1.7  & 3.9$\pm$1.6 \\
\makebox[6em][l]{NGC~507} ($r>9'$)    $\dotfill$ & 3.7$\pm$2.7  & 3.3$\pm$2.4 \\
Galactic average \citep{mccammon02}   $\dotfill$ & 4.8$\pm$0.8 & 1.6$\pm$0.4 \\
\hline 
\end{tabular}
\end{center}
\end{table*}

\begin{figure}
\begin{center}
\begin{minipage}{0.45\textwidth}
\centerline{
\FigureFile(\textwidth,\textwidth){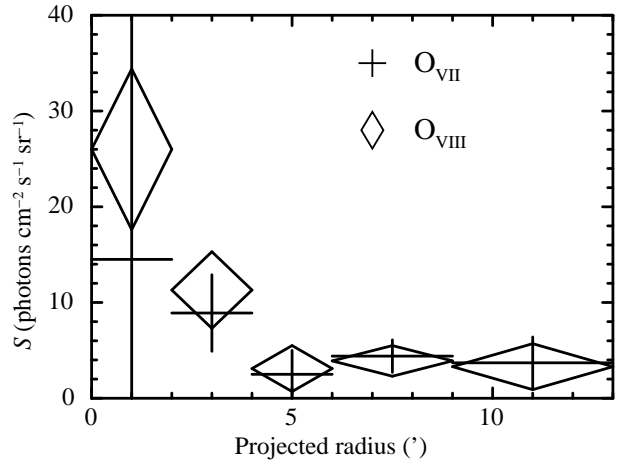}}
\caption{
Intensities of O\emissiontype{VII} and O\emissiontype{VIII} lines
at each annulus of NGC~507 in units of photons~cm$^{-2}$~s$^{-1}$~sr$^{-1}$.
Numerical values are shown in table~\ref{tab:7}.
}\label{fig:7}
\end{minipage}
\end{center}
\end{figure}

We examined the surface brightness of the O\emissiontype{VII} and
O\emissiontype{VIII} emission lines in order to look into the origin of the O
lines directly from the line intensities.  The surface
brightness of the lines
was derived by fitting the annular spectrum with a {\it power-law} +
{\it gaussian} + {\it gaussian} model.  In this fit, we fixed the
Gaussian $\sigma$ to be 0, and allowed the energy center of the two
Gaussians to vary within 555--573~eV or 648--658~eV for
O\emissiontype{VII} or O\emissiontype{VIII} line, respectively.  The
derived line intensities are summarized in table~\ref{tab:7} and
figure~\ref{fig:7}.  There is a clear excess of the
O\emissiontype{VIII} intensity towards the cluster center, while
O\emissiontype{VII} one is consistent to be constant.  This is a clear
evidence that the O\emissiontype{VIII} line is associated with the
ICM itself, on the other hand, O\emissiontype{VII} may be due mainly
to the Galactic origin.

\section{Discussion}\label{sec:discuss}

\subsection{Metallicity Distribution in ICM}

\begin{figure}
\centerline{\FigureFile(0.45\textwidth,8cm){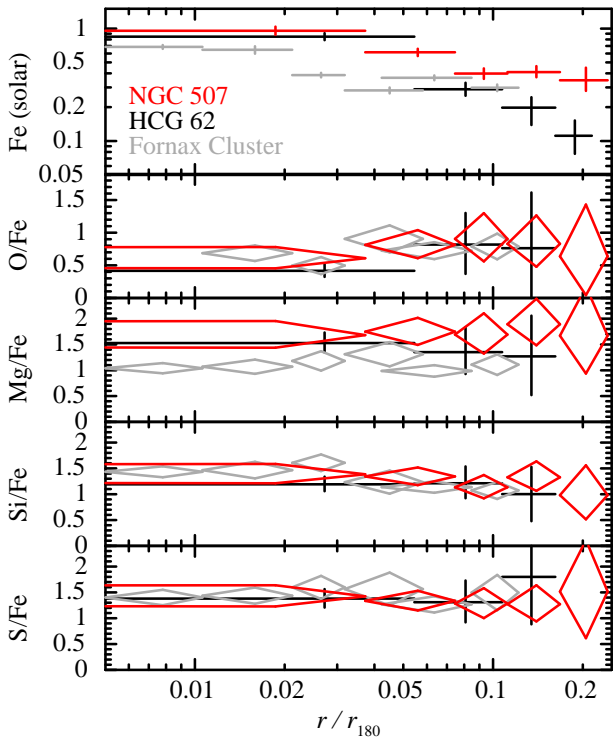}}
\vspace*{-1ex}
\caption{ Comparison of the abundance ratio for NGC~507 (red) with
results for the HCG~62 group (black; \cite{tokoi08}) and the Fornax
cluster (gray; \cite{matsushita07b}). Fe abundance and the O/Fe,
Mg/Fe, Si/Fe, and S/Fe abundance ratios in solar units
\citep{anders89} are plotted against the projected radius scaled by
the virial radius, $r_{180}$, in all panels.  }\label{fig:8}
\end{figure}

The present Suzaku observation of NGC~507 showed abundance
distribution of O, Mg, Si, S, and Fe out to a radius of $13'\simeq
260$~kpc as shown in figure \ref{fig:5}. Ne abundance has large
ambiguity due to the strong coupling with Fe-L lines.  Distributions
of Mg, Si, S, and Fe are quite similar to each other, while O profile
in the outer region has a large uncertainty.  We plotted abundance
ratios of O, Mg, Si, and S over Fe as a function of the projected
radius in figure~\ref{fig:8}.  The ratios Mg/Fe, Si/Fe and S/Fe are
consistent to be a constant value around 1.5--2, while O/Fe ratio for
the innermost region ($r<2'$) is significantly lower around 0.6.  In
addition, the O/Fe ratio suggests some increase with radius.

Recent Suzaku observations have presented abundance profiles in
several other systems: an elliptical galaxy NGC~720 \citep{tawara08},
the Fornax cluster and NGC~1404 \citep{matsushita07b}, and a cluster
of galaxies Abell~1060 \citep{sato07a} and AWM~7 \citep{sato08}.  All
systems show very similar value of Si/Fe ratio, to be 1--1.5.  Mg/Fe
ratio is slightly higher in HCG~62, Abell ~1060 and AWM~7 than in
other systems, i.e.\ NGC~720, Fornax cluster, and NGC~1404\@.  We
compare metal abundances of NGC~507 with those in the Fornax cluster
and HCG~62 as shown in figure~\ref{fig:8}.  The solar abundance by
\citet{feldman92}, giving $\rm [Fe/H] = 3.24\times 10^{-5}$ and
employed by \citet{matsushita07b}, was scaled by a factor of 0.7 to
match the \citet{anders89} value.  The Fe abundance of NGC~507 is
almost the same as that of HCG~62 in the central region, while the
Fornax cluster shows a lower value.  Abundance ratios of O/Fe, Mg/Fe,
Si/Fe, and S/Fe are quite similar for these 3 systems at $r\sim 0.1\;
r_{180}$. 
Therefore, the abundance ratios show closer similarity than the
absolute abundance values among different systems. The efficiency of
the metal enrichment may depend on parameters such as age,
starformation efficiency, contribution from cD galaxies. However, the
relative contribution of SNe Ia and II and the process of metal
mixing in the ICM seem to be quite similar for different clusters and
groups.

\citet{tamura04} reported abundance ratios for 19 clusters studied
with XMM-Newton, and the mean Si/Fe ratio in cool and medium
temperature clusters with $kT < 6$~keV was $\sim 1.4$.  This is
consistent with the Suzaku results for groups and poor clusters
including NGC~507.  Their O/Fe ratio, $\sim0.6$, in the cluster core
also agrees with the Suzaku results including our NGC~507 case.
\citet{matsushita03,matsushita07a} also reported abundance ratio for
M87 and the Centaurus cluster, respectively, based on XMM-Newton
observations.  M87 showed Mg/O ratio to be $\sim 1.3$ in the central
region, and the Centaurus cluster indicated O/Fe and Si/Fe ratios
within $8'$ to be consistent with our results.  The Si/Fe ratio for
NGC~507 with XMM-Newton \citep{kim04} and Chandra \citep{humphrey06}
is $\sim 1$, which is almost the same as our result.

\citet{sato07b} studied contributions of SNe Ia and II 
to the metal enrichment, based on Suzaku results of NGC~507,
Abell~1060, AWM~7, and HCG~62 \citep{sato07a,sato08,tokoi08}.
\citet{sato07b} showed the number ratios of SNe II to Ia 
to be $\sim3.5$ for the ICM of the above systems. These results also 
suggest that the clusters and groups have passed the same 
metal enrichment process in the ICM.

\subsection{Metal Mass-to-Light Ratio}

\begin{figure*}
\centerline{
\FigureFile(0.45\textwidth,1cm){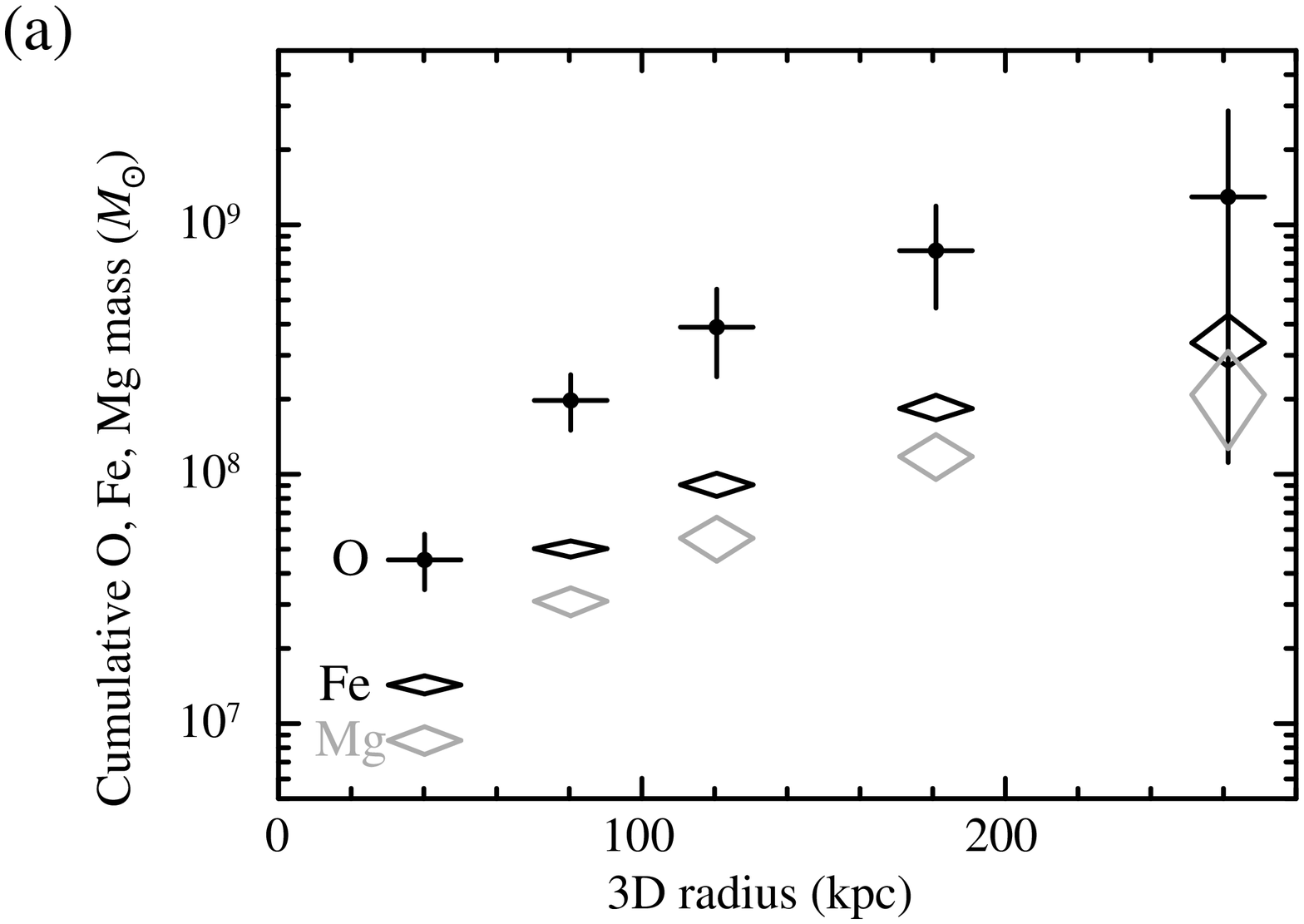}%
\hspace*{0.05\textwidth}
\FigureFile(0.45\textwidth,1cm){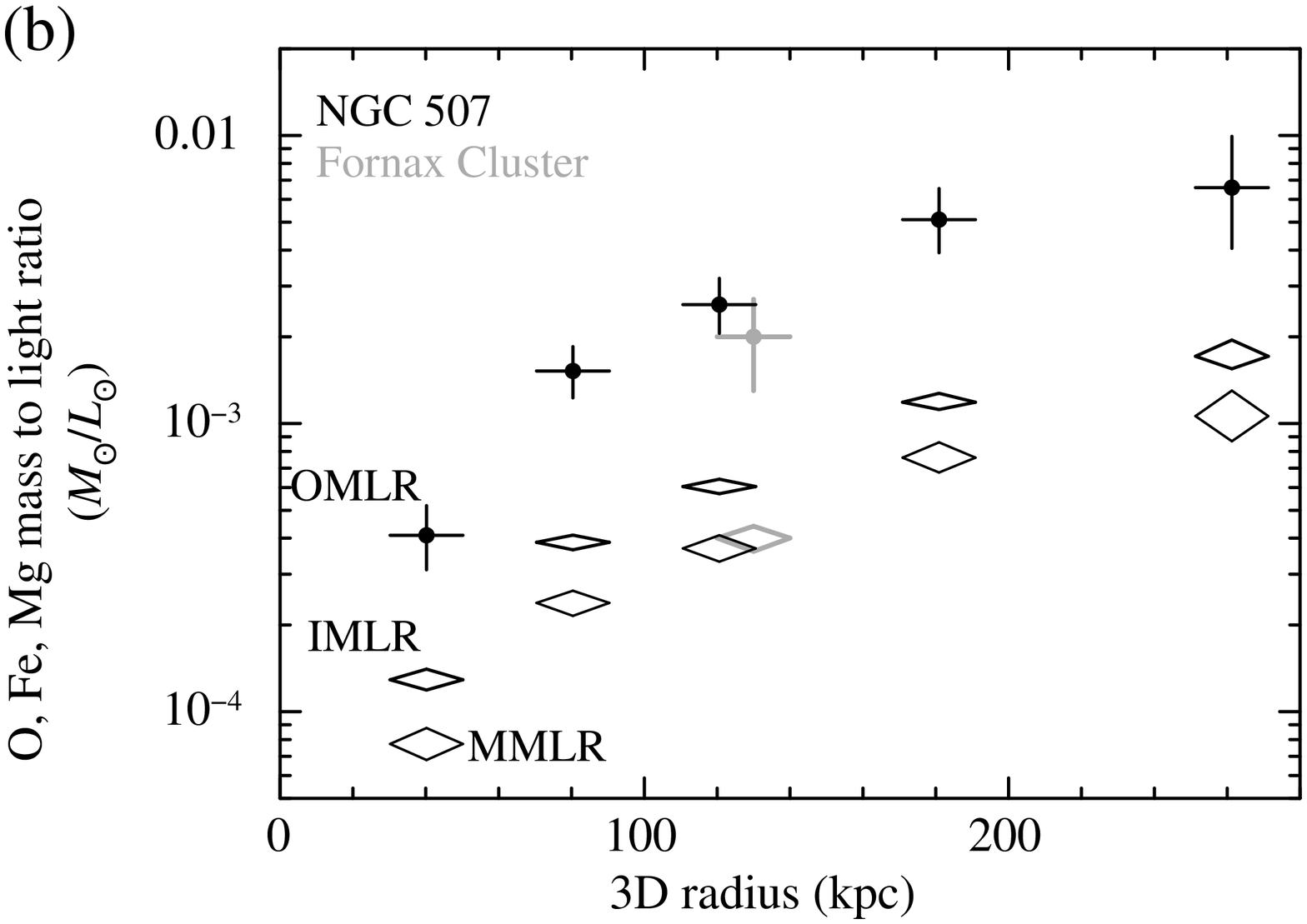}%
}
\caption{
(a) Cumulative mass, $M(<R)$, within the 3-dimensional radius, $R$,
for O, Fe, and Mg in NGC~507,
based on the combination of the abundance determination with Suzaku
and the gas mass profile with XMM-Newton.
(b) Ratio of the O, Fe, and Mg mass in units of $M_\odot$
to the $B$ band optical luminosity in units of $L_\odot$
(OMLR, IMLR, and MMLR, respectively) against
the 3-dimensional radius. Black symbols represent our results
for NGC~507, and gray symbols of OMLR and IMLR are
for the Fornax cluster by \citet{matsushita07b}.
}\label{fig:9}
\end{figure*}

\begin{table*}
\caption{Comparison of IMLR, OMLR and MMLR for all systems.
}\label{tab:8}
\begin{center}
\begin{tabular}{lrrcrrl}
\hline\hline
& IMLR & OMLR & MMLR & \multicolumn{1}{c}{$r$ (kpc/$r_{180}$)} & \multicolumn{1}{c}{$k\langle T \rangle$} & Reference \\
\hline
Suzaku & & & & \\
NGC~720 $\dotfill$   & $1\times 10^{-4}$ & $4\times 10^{-4}$ & -- &25/0.04 & $\sim 0.56$ keV &       \citet{tawara08} \\
Fornax $\dotfill$  & $4\times 10^{-4}$ & $2\times 10^{-3}$ & -- &130/0.13 & $\sim 1.3$ keV &         \citet{matsushita07b} \\
NGC~507 $\dotfill$  & $1.7\times 10^{-3}$ & $6.6\times 10^{-3}$ &$1.1\times 10^{-3}$  &260/0.24 & $\sim 1.5$ keV &      This work\\
HCG~62 $\dotfill$  & $4.6\times 10^{-3}$ & $3.8\times 10^{-2}$ &$1.5\times 10^{-3}$  &230/0.21 & $\sim 1.5$ keV &       \citet{tokoi08} \\
A~1060 $\dotfill$   & $4.0\times 10^{-3}$ & $4.3\times 10^{-2}$ &$1.6\times 10^{-3}$  &380/0.25 & $\sim 3$ keV &        \citet{sato07a} \\
AWM~7 $\dotfill$   & $1.1\times 10^{-2}$ & $5.8\times 10^{-2}$ &$9.4\times 10^{-3}$  & 570/0.35 & $\sim 3.5$ keV&       \citet{sato08}\\
\hline
XMM-Newton & & & & \\
Centaurus $\dotfill$ & $4\times 10^{-3}$ & $3\times 10^{-2}$ &-- &190/0.11 & $\sim 4$ keV &        \citet{matsushita07a} \\
\hline
\end{tabular}
\end{center}
\end{table*}

We examined the parameter of metal mass-to-light ratios for oxygen,
iron, and magnesium (OMLR, IMLR, and MMLR, respectively) to compare
the ICM metal distribution with the stellar mass profile.  The metal
mass profiles shown in figure~\ref{fig:9}(a) were calculated from the
3-dimensional gas mass profile, which is based on the surface
brightness profile by XMM-Newton (see also figure~\ref{fig:2} and
subsection \label{subsec:arf}), the gas density in the central region,
$r<2'$, through our spectral fit as shown in table~\ref{tab:5}, and
the abundance profile measured with Suzaku.  The derived iron, oxygen,
and magnesium mass within the 3-dimensional radius of $r< 260$~kpc are
$3.4\times 10^8$, $1.3\times 10^9$, and $2.1\times 10^8$ $M_\odot$,
respectively.

Historically, B-band luminosity has been used for the estimation of the
stellar mass \citep{makishima01}. We calculated the B-band luminosity
in NGC507 by translation from the Two Micron All Sky Survey (2MASS)\footnote{
The database address: {\tt http://www.ipac.caltech.edu/2mass/}} with
an appropriate color $B-K=4.2$ for early-type galaxies in
\citet{lin04}, along with the Galactic extinction, $A_B=0.267$, 
from NASA/IPAC Extragalactic Database (NED) in the direction of NGC~507\@.
We used $2^{\circ}\times2^{\circ}$ data set centered at the NGC~507
coordinate as shown in table 1, and subtracted the $r>1^{\circ}$ region 
as a background, and also removed the NGC~499 region. In addition, 
we deprojected the luminosities along with the annuli of the spectral
fits. The resultant luminosity within this Suzaku observation, $r<13'$, 
is $2\times10^{11}~L_{\odot}$ in B-band. The radial luminosity 
profile corresponds to the value of figure 9 (a) divided by (b).

We calculated the integrated values of OMLR, IMLR, and MMLR within
$r\lesssim 260$~kpc as shown in figure~\ref{fig:9}(b), and their
values turned out to be $\sim 6.6\times 10^{-3}$, $\sim 1.7\times
10^{-3}$, and $\sim 1.1\times 10^{-3}$ $M_{\odot}/L_{\odot}$,
respectively.  The errors are only based on the statistical errors of
metal abundance in the spectral fit, and the uncertainties of the gas
mass profile and the luminosity of member galaxies are not
included.  The IMLR values are consistent with the collective
results with ASCA by \citet{makishima01}.  The MMLR and IMLR show
similar steep increase with radius up to $r \sim 150$~kpc and seem to
reach almost a plateau at 150--300~kpc.  This tendency is similar to the
IMLR profile with XMM-Newton for M~87 and the Centaurus cluster by
\citet{matsushita07a} within $r\lesssim 0.1~r_{180}$.

We also note that the derived OMLR and IMLR curves for NGC~507 in
figure~\ref{fig:9} are very similar to those for the Fornax cluster 
\citep{matsushita07b} as seen in figure~\ref{fig:9}(b).  
We summarize recent measurements of
IMLR and OMLR in table~\ref{tab:8}, where one can compare the present
results with those in Fornax cluster, Centaurus cluster and NGC~720.
It is suggested that smaller systems with lower gas temperature tend 
to show not only lower IMLR in \citet{makishima01} but also lower OMLR 
and MMLR, even though the scatter of the data is large.
Because O and Mg are mainly synthesized by SNe II, while Fe is 
by both SNe Ia and II, as shown in \citet{sato07b},
OMLR and MMLR show the ratios of heavy stars to galaxies. 
Although our observation within $r\lesssim 0.3~r_{180}$ show no clear
difference between the distribution of IMLR and OMLR and MMLR, we need 
to observe the outer region to the virial radius in order to 
know the metal enrichment process of the ICM.

\subsection{Discontinuity in the Surface brightness}

\begin{figure}
\begin{center}
\begin{minipage}{0.45\textwidth}
\centerline{
\FigureFile(\textwidth,\textwidth){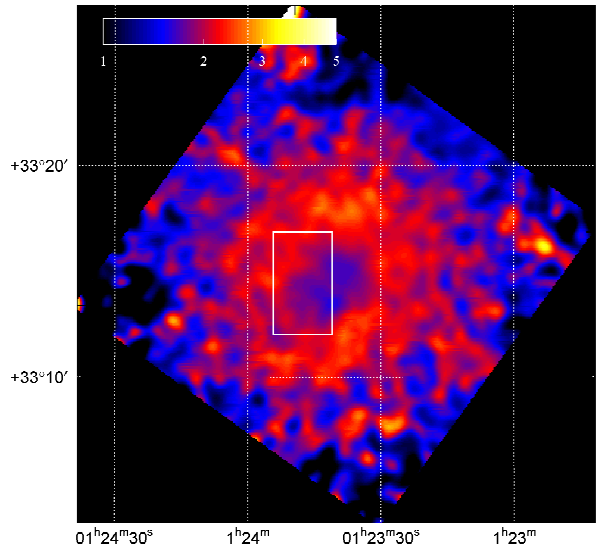}}
\caption{ Hardness ratio image for the two energy bands, 1--2 keV to
0.5--1 keV, in the central region of the Suzaku observation.  The
observed images by XIS0-3 were superposed on the sky coordinate after
removing calibration source region, and smoothed with a Gaussian with
$\sigma=16$ pixel $\simeq 17''$.  The white box corresponds to the
X-ray discontinuity recognized by \citet{kraft04} }\label{fig:10}
\end{minipage}
\end{center}
\end{figure}

Since previous X-ray observations suggested possible dynamical
features in the NGC507 system (e.g.\ cite{kraft04}), we looked into
temperature structures which are out of axis symmetry in the form of
2-dimensional map of hardness ratio.  Figure~\ref{fig:10} shows the
hardness ratio image based on the intensities in 1--2 keV and 0.5--1
keV\@.  Although a vignetting correction was not performed, the two
energy bands show very similar vignetting features.  The transmission
drop due to the OBF contamination was not corrected for either, which
is lager for the 0.5--1 keV band and at the central part of the CCD
image. Therefore, the hardness ratio should show a systematic drop
from the center to the outer regions even for a constant temperature
ICM\@.  In the hardness image, the central region shows clearly lower
hardness, and these feature is reported for HCG~62 observation in
\citet{tokoi08}.  Including this feature, deviation from the circular
symmetry is not significantly recognized in the hardness ratio image
of NGC507\@.

\citet{kraft04} found a sharp edge or discontinuity in the X-ray
surface brightness by Chandra in the east and south east region at
$\sim2'$ from the center.  The discontinuity corresponds to the radius
where the influence of the central cool component becomes weak, as
indicated in figure~\ref{fig:10}.  \citet{kraft04} suggested three
possibilities for the discontinuity; 1) the edge is caused by a motion
of NGC~507 with respect to the lager scale dark matter potential of
the group, 2) the pressure across the boundary is balanced by an
unseen relic radio lobe produced in an earlier epoch of nuclear
activity, 3) the discontinuity is created by the subsonic/transonic
expansion of the relatively weak lobe currently observed lying
interior to the discontinuity.  Regarding the possibility 1,
\citet{kraft04} noted that the temperatures inside and outside the
discontinuity are almost the same.  However, while \citet{kraft04}
fitted with a single temperature model, we employed the two
temperature model.  As shown in figure~\ref{fig:6} and \ref{fig:10},
relative intensity of the cool component drops fairly sharply with
radius around the discontinuity region.

\bigskip
Authors thank the referee for providing valuable comments. 
Part of this work was financially supported by the Ministry of
Education, Culture, Sports, Science and Technology of Japan,
Grant-in-Aid for Scientific Research
Nos.\ 14079103, 15340088, 15001002, 16340077, 18740011, 19840043.

\end{document}